  \documentclass[traditabstract]{aa}

\usepackage{graphicx}
\usepackage{natbib}
\usepackage{lscape}
\usepackage{supertabular}
\usepackage{xcolor}

\newcommand{\vsini}{$v \sin{i}$}
\newcommand{\hei}[0]{\ion{He}{i}}
\newcommand{\heir}[0]{\hei\,IR}
\newcommand{\cab}[0]{\ion{Ca}{ii}~IRT$_1$}

\usepackage{txfonts}

\begin{document}

\title{The CARMENES search for exoplanets around M dwarfs}
\subtitle{Variability of the \hei\, line at 10830 \AA}\thanks{Full Table 2 is only available in 
electronic form
at the CDS via anonymous ftp to cdsarc.u-strasbg.fr (130.79.128.5)
or via http://cdsweb.u-strasbg.fr/cgi-bin/qcat?J/A+A/}

\author{B. Fuhrmeister\inst{\ref{inst1}}, S. Czesla\inst{\ref{inst1}},  L. Hildebrandt\inst{\ref{inst1}}, E. Nagel\inst{\ref{inst1},\ref{inst13}}, J. H. M. M. Schmitt\inst{\ref{inst1}}
  \and  S.~V.~Jeffers\inst{\ref{inst2}}
  \and  J.~A.~Caballero\inst{\ref{inst3}}
  \and  D.~Hintz\inst{\ref{inst1}}
  \and  E.~N.~Johnson\inst{\ref{inst2}}
  \and  P.~Sch\"ofer\inst{\ref{inst2}}
  \and M.~Zechmeister\inst{\ref{inst2}}
  \and   A.~Reiners\inst{\ref{inst2}}
  \and   I.~Ribas\inst{\ref{inst4},\ref{inst5}}
  \and P.~J.~Amado\inst{\ref{inst6}}
  \and  A.~Quirrenbach\inst{\ref{inst7}}
  \and  L.~Nortmann\inst{\ref{inst2}}
  \and  F.~F.~Bauer\inst{\ref{inst6}}
  \and  V.~J.~S.~B\'ejar\inst{\ref{inst8},\ref{inst9}}
  \and  M.~Cort\'es-Contreras\inst{\ref{inst3}}
  \and  S.~Dreizler\inst{\ref{inst2}}
  \and  D.~Galad\'{\i}-Enr\'{\i}quez\inst{\ref{inst12}}
  \and  A.~P.~Hatzes\inst{\ref{inst13}}
  \and  A. Kaminski\inst{\ref{inst7}}
  \and  M.~K\"urster\inst{\ref{inst14}}
  \and  M.~Lafarga\inst{\ref{inst4},\ref{inst5}}
  \and  D.~Montes\inst{\ref{inst10}}}

\institute{Hamburger Sternwarte, Universit\"at Hamburg, Gojenbergsweg 112, D-21029 Hamburg, Germany\\
  \email{bfuhrmeister@hs.uni-hamburg.de}\label{inst1}
        \and
        Institut f\"ur Astrophysik, Friedrich-Hund-Platz 1, D-37077 G\"ottingen, Germany\label{inst2} 
        \and
        Centro de Astrobiolog\'{\i}a (CSIC-INTA), ESAC, Camino Bajo del Castillo s/n, E-28692 Villanueva de la Ca\~nada, Madrid, Spain \label{inst3}
        \and
        Institut de Ci\`encies de l'Espai (CSIC), Campus UAB, c/ de Can Magrans s/n, E-08193 Bellaterra, Barcelona, Spain\label{inst4}
        \and
        Institut d'Estudis Espacials de Catalunya, E-08034 Barcelona, Spain\label{inst5}
           \and 
        Instituto de Astrof\'isica de Andaluc\'ia (CSIC), Glorieta de la Astronom\'ia s/n, E-18008 Granada, Spain\label{inst6} 
        \and 
        Landessternwarte, Zentrum f\"ur Astronomie der Universit\"at Heidelberg, K\"onigstuhl 12, D-69117 Heidelberg, Germany\label{inst7} 
        \and
        Instituto de Astrof\'{\i}sica de Canarias, c/ V\'{\i}a L\'actea s/n, E-38205 La Laguna, Tenerife, Spain\label{inst8}
        \and
        Departamento de Astrof\'{\i}sica, Universidad de La Laguna, E-38206 Tenerife, Spain\label{inst9} 
        \and
        Facultad de Ciencias F\'{\i}sicas, Departamento de F\'{\i}sica de la Tierra y Astrof\'{\i}sica; IPARCOS-UCM (Instituto de F\'{\i}sica de Part\'{\i}culas y del Cosmos de la UCM), Universidad Complutense de Madrid, E-28040 Madrid, Spain\label{inst10} 
        \and
        Departamento de Explotaci\'on y Prospecc\'on de Minas, Escuela de Minas, Energ\'{\i}a y Materiales, Universidad de Oviedo, E-33003 Oviedo, Asturias, Spain\label{inst11}
        \and
        Centro Astron\'omico Hispano-Alem\'an (MPG-CSIC), Observatorio Astron\'omico de Calar Alto, Sierra de los Filabres, E-04550 G\'ergal, Almer\'{\i}a, Spain\label{inst12} 
        \and
        Th\"uringer Landessternwarte Tautenburg, Sternwarte 5, D-07778 Tautenburg, Germany\label{inst13} 
        \and
        Max-Planck-Institut f\"ur Astronomie, K\"onigstuhl 17, D-69117 Heidelberg, Germany\label{inst14} 
       }
        
\date{Received dd/05/2020; accepted dd/mm/2020}

\abstract
    {The \hei\, infrared (IR) triplet at 10830 \AA\, is 
      known as an activity indicator in solar-type stars and
      has become a primary diagnostic in exoplanetary transmission spectroscopy.
       \heir\, lines are a tracer of the stellar extreme-ultraviolet irradiation
      from the transition region and corona.
      We study the variability of the \hei\ triplet lines in
      a spectral time series of 319 M~dwarf stars that was obtained with the CARMENES high-resolution optical and
      near-infrared spectrograph at Calar Alto.
      We detect \heir\ line variability in 18\,\% of our sample stars,
      all of which show H$\alpha$ in emission. Therefore, we find detectable
      \hei\ variability in 78\,\% of the sub-sample of stars with H$\alpha$ emission.
      Detectable variability is strongly concentrated in the latest spectral sub-types, where
      the \hei\ lines during quiescence are typically weak.
      The fraction of stars with detectable \hei\ variation remains lower than 10\,\% for stars
      earlier than M3.0\,V, while it exceeds $30$\,\% for the later spectral sub-types.
      Flares are accompanied by particularly pronounced line variations, including strongly broadened
      lines with red and blue asymmetries. However, we also find evidence for enhanced \hei\ absorption, which is 
      potentially associated with increased high-energy irradiation levels
      at flare onset. 
      Generally, \hei\ and H$\alpha$ line variations tend to be correlated, with H$\alpha$ being
      the most sensitive indicator in terms of pseudo-equivalent width variation. This makes
      the \hei\ triplet a favourable target for planetary transmission spectroscopy.
%
      }
    
\titlerunning{Variability in the \ion{He}{i} 10830 \AA\, line}
\authorrunning{B. Fuhrmeister et~al.}
\maketitle


\section{Introduction}

Late-type stars are known to show ``activity'', which is a term that summarises a zoo of phenomena fuelled by energy from
the stellar magnetic field. In particular, M~dwarfs frequently exhibit strong activity as evidenced,
for example, by coronal X-ray radiation \citep[e.\,g.][]{Pizzolato,robrade},
prominent chromospheric emission lines \citep{Gizis2000,Walkowicz2009,Houdebine2012,Kowalski2017},
variable temperature-sensitive molecular bands \citep{Patrick},
or high levels of photometric modulation \citep{Rebull2016,SM16,DA19}.

 As a
result of new observational possibilities, one chromospheric activity tracer,
the \hei\  infrared (IR)  triplet at 10830 \AA,\, has recently become
 a highly promising diagnostic of the outer
atmospheres of exoplanets and their dynamics \citep[e.g.][]{Spake2018, Salz2018, Nortmann2018,
Allart2018, Mansfield2018, Alonso-Floriano2019}. This  IR triplet is the only known tracer
of planetary mass loss observable from the ground.
Planetary transmission spectroscopy is based on the analysis of changes in the profiles
and depths of the observed spectral lines.
Effects attributable to
activity-related temporal variability or
an inhomogeneous line profile distribution across the stellar disc, such as those
produced by active regions, are major nuisances in the study of exoplanetary atmospheres.
Studying the \hei\ transmission signal of HD~189733\,b, \citet{Salz2018}
estimate that a pseudo-signal introduced by stellar activity of the K-dwarf host star HD~189733
might be responsible for up to 80\,\% of their measured signal. On the other hand, \citet{Cauley2018} used  simulations to find that the \heir\ line should be less affected by contamination caused by stellar
activity than other chromospheric lines.
Nevertheless, knowledge of stellar line variability is a crucial ingredient in  studies
of exoplanetary spectra.

Activity-induced variability occurs on very different timescales. 
Short-term variability
can be caused by transient events such as flares, which are frequently observed on M~dwarfs 
both in photometric light curves \citep[e.g.][]{Walkowicz2011,Hawley2014,Yang2017,Doyle2018} and spectroscopic observations
\citep{ADLeoflare,CNLeoflare,Schmidt2012}.
The flare frequency increases along the M-dwarf sequence with flare duty cycles reaching 3\,\% for late-type M~dwarfs 
\citep{Hilton2010}. On the rotational timescale, ramifications of 
photospheric spots and chromospheric plages can be observed when they rotate across 
the visible hemisphere \citep{Barnes,SM2017,Newton2017,DA19}. 
Finally, activity cycles can impose long-term modulation on the scale of years \citep{Baliunas,Berdyugina2005,SM16}.

To study activity spectroscopically  activity-sensitive chromospheric emission lines are used,
the best known are the \ion{Ca}{ii} H \& K lines at $3968$~\AA\, and $3934$~\AA\, \citep{Baliunas,Wright2004}.
Since M~dwarfs are rather faint in the blue wavelength range, studies of these lines require a
substantial observational effort \citep{Walkowicz2009}.
Additionally, \ion{Ca}{ii} IR triplet lines at around 8500 to 8600~\AA\, can
be used as variability tracers,
although they are not as sensitive as H\&K lines \citep{Johannes,Jeffers2018}.
The same applies to the \hei\ D$_{3}$ line at $5876$~\AA\ and the
neighbouring \ion{Na}{i} D doublet lines, which can moreover be affected by airglow \citep{airglow}.
As a consequence, many activity studies of M dwarfs have focussed on the very sensitive and accessible H$\alpha$ line, which
however may not sample the same phenomena as \ion{Ca}{ii} H \& K lines \citep{Walkowicz2009,Jeffers2018,Patrick}.

Prominent activity-sensitive lines in the near-infrared (NIR) regime are the \hei\ IR triplet lines at
(vacuum) wavelengths of 10\,832.057, 10\,833.217, and 10\,833.306~\AA.
This triplet is formed by transitions between the meta-stable $2^{3}\mathrm{S}$
and the $2^{3}\mathrm{P}$ levels. The two reddest components dominate the triplet and remain
unresolved in the majority of studies, which is
why we refer to this complex as the \hei\ IR line for short.
While the \hei\ line has long been studied in the Sun \citep[e.\,g.][]{Zirin1968, Zirin1982},
it is a relatively new addition in the context of stellar observations \citep{Sanz-Forcada2008, Andretta2017}; these observations have only become feasible after the advent of suitable high-resolution infrared spectrographs.
As the meta-stable $2^{3}\mathrm{S}$ state of neutral helium has an excitation potential of about $20$~eV,
it is thought to be populated by photo-ionisation--recombination processes triggered by
extreme-ultraviolet (EUV) and X-ray emission from the stellar transition region and corona
with wavelengths below the helium ionisation edge at $504$~\AA\, \citep{Zirin1988}.
This makes the \hei\ line a prominent tracer of the stellar EUV emission, which is
typically heavily affected by interstellar absorption
and not directly observable with ground-based instrumentation.

A highly stabilised spectrograph, CARMENES  simultaneously covers the optical and 
near-infrared range, including the \hei\ IR line region, at high resolution (Sect.~\ref{sec:obs}).
This has made it a prime instrument to study both planetary and stellar \hei\ lines.
In \citet{hepaper}, we present
a systematic study of the \hei\ IR line in M~dwarfs with a focus on the
time-averaged properties of the line. Our study shows that the line is
usually a strong feature in early M~dwarfs. Yet, its pseudo-equivalent width (pEW) decreases towards
later spectral sub-types becoming undetectable around sub-type M5.0\,V (see also Sect. \ref{sec:generalvar}).
We revisit the spectral time series of the 319 stars also considered in \citet{hepaper} and
extend our previous analysis into the time domain; we also examine the variability amplitude of the \hei\ IR line,
the relation to other activity indicators, and its behaviour during flares.
We intend to interpret this in terms of the reliability of \heir\ line detection and analysis
of exoplanet
atmospheres \citep{Seager2000}. 

Our paper is structured as follows: in Sect. 2 we give an overview of the used
data and our method for pEW measurements. Subsequently we describe our findings on variability in
the \hei\ IR line in Sect. 3, while in Sect. 4 we concentrate on some outstanding examples and the conclusions
we draw thereof. In Sect. 5 we point out the implications for exoplanet atmosphere studies
analysing the \heir\ line. Finally, in Sect. 6, we summarise our findings.

\section{Observations and measurement method}
\label{sec:obs}
\subsection{Used data and their reduction}
All spectra used for the present analysis were taken
with the CARMENES spectrograph, mounted at the 3.5\,m Calar Alto 
Telescope \citep{CARMENES1}.
The CARMENES instrument is a highly stabilised spectrograph covering the wavelength range
from 5200 to 9600\AA\, in the visual channel (VIS) and from 9600\,\AA\,
to 17\,100\,\AA\, in the NIR
channel. The instrument  provides a spectral resolution of
$\sim$ 94\,600 in VIS and $\sim$ 80\,400 in NIR. 
The CARMENES consortium is conducting a 750-night 
survey, targeting $\sim$350 M~dwarfs to find low-mass exoplanets \citep{AF15a, Reiners2017}.
To date, CARMENES has obtained more than 16\,000 high-resolution visible and NIR spectra.
Since the cadence of the spectra is optimised for the planet search, as a rule
no continuous time series are obtained and the typical observing frequency is at maximum once per night.

The spectra of the monitored 319 stars were reduced using the CARMENES reduction pipeline
\citep{pipeline}. Subsequently, they were corrected for barycentric and other radial velocity motions, as well as for secular
acceleration and telluric absorption \citep{Evangelos} 
using the {\tt molecfit} package\footnote{\tt{https://www.eso.org/sci/software/pipelines/skytools/molecfit}}.  
The \hei\ IR line region is contaminated with
OH airglow lines. The four relevant lines belong to two so-called
$\Lambda$-doublets at
wavelengths of 10\,834.241 and 10\,834.338 \AA\ and 10\,832.412 and
10832.103 \AA\
\citep[][]{Phillips2004, Oliva2015}. The redder doublet remains unresolved
and is
the stronger one. All lines correspond to transitions between levels with
vibrational
quantum numbers of five and two and the same rotational quantum number,
which suggests a
relation between their intensities. We used around $1600$ sky-fibre spectra
obtained by CARMENES to verify this hypothesis and determined the intensity
ratio between
the doublets by fitting Gaussian components to the OH sky spectra. As the
redder doublet tends to be prominent in the science spectrum, we used it to constrain
the normalisation of
the thus determined OH model with fixed intensity ratios and subtracted it
from the science spectrum.
This procedure led to a substantial reduction of the telluric line
contamination. Naturally,
however, some spectra remain affected by varying degrees of correction
artefacts.


\subsection{Equivalent width measurement}\label{sec:equivwidth}

We measured the pEW of the \hei\, IR line, the
H$\alpha$ line, the bluest component of the
\ion{Ca}{ii} infrared triplet lines (\cab\ line), and the \ion{He}{i}
D$_{3}$ line.
For the \hei\, IR line we treated the blended two reddest components as one
single line at
10\,833.26~\AA. The precise integration and reference bands are given in
Table~\ref{ew}.
Thus, a time series of pEW measurements was obtained for each star.
From these pEW time series, we calculated the mean pEW
and the median average deviation about the median (MAD), which is
a robust estimator
of the scatter \citep[][]{Hampel1974, Rousseeuw1993, Czesla2018}.
For a Gaussian distribution, the MAD multiplied by $1.4826$ is an estimator
of the standard deviation.
We decided not to use the mean average deviation of the mean because it is
more severely
influenced by outliers that are mainly caused by artefacts such as cosmic
rays and leftovers
from the telluric correction in the case of the \hei\, IR line. The error of
the
pEW depends strongly on the signal-to-noise ratio of the spectra and is,
therefore, much lower
for a bright early M dwarf than for a fainter mid-type M dwarf because the
exposure
times never exceed 1800 s. Moreover, the pEW(\hei) changes if there
are artefacts included in the integration interval.

Finally, we quantified the degree of correlation between the pEW time series
of the
\heir\ line and the H$\alpha$, \ion{He}{i} D$_{3}$,
and \cab\ lines by
computing Pearson's correlation coefficient $r$ and their
p-values. For comparison, we also computed Pearson's correlation coefficient
for the pEW time series of the H$\alpha$ and \cab\ lines.
In Table~\ref{measurements}, we list
the values for all stars identified by their CARMENES identification
(Karmn); the full tables are available at CDS.

\subsubsection{Error analysis}
\begin{figure}
\begin{center}
\includegraphics[width=0.5\textwidth, clip]{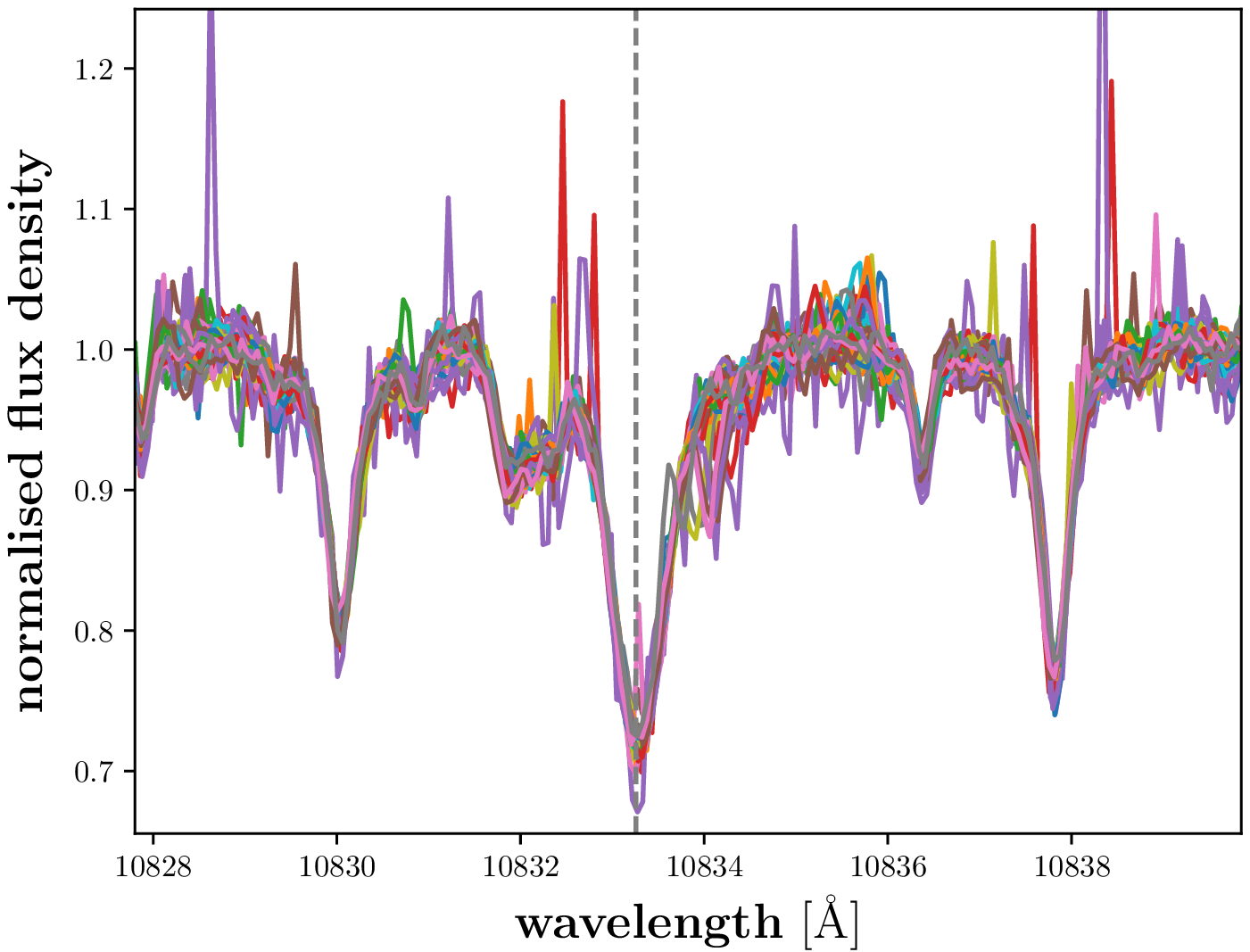}\\
\includegraphics[width=0.5\textwidth, clip]{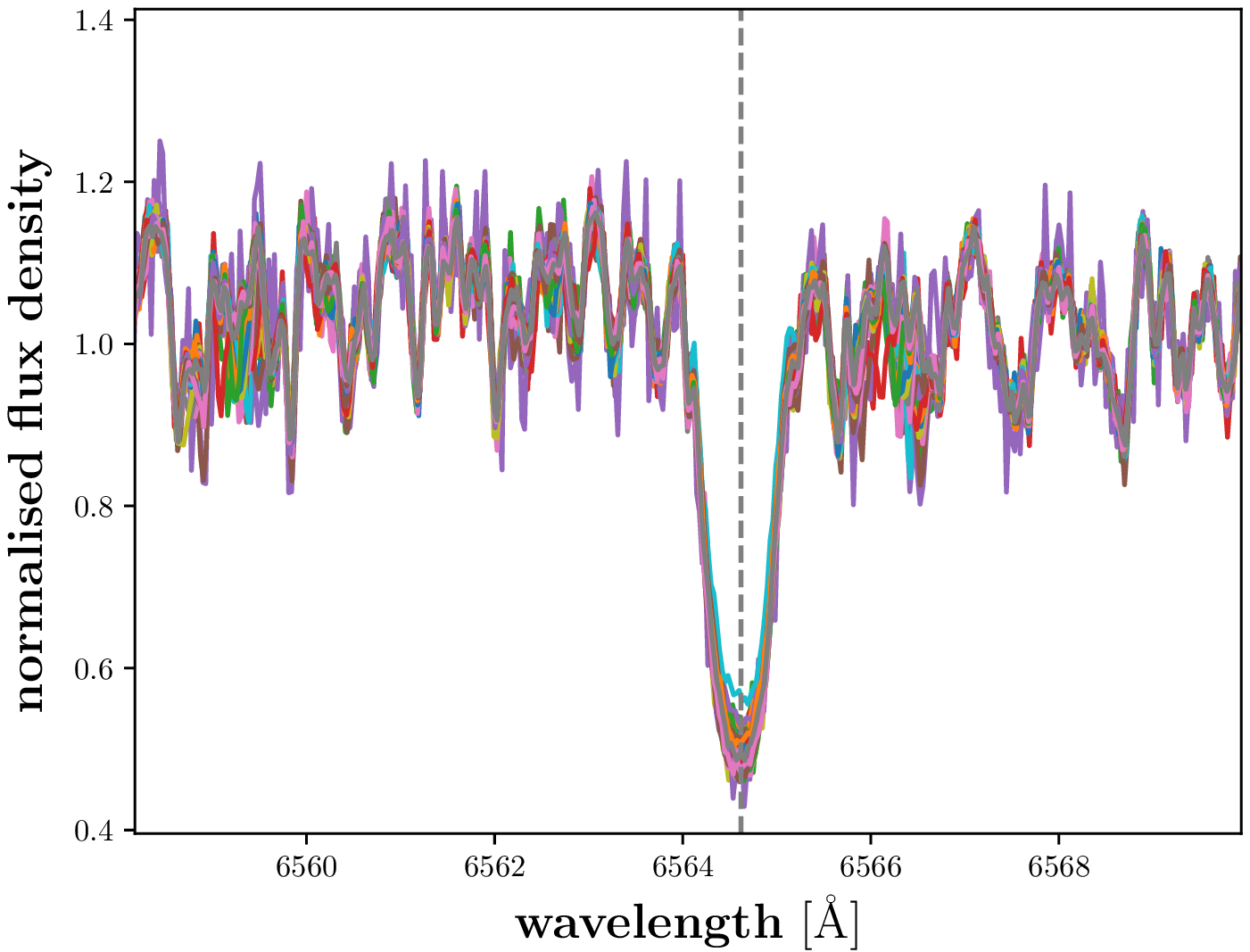}\\
\includegraphics[width=0.5\textwidth, clip]{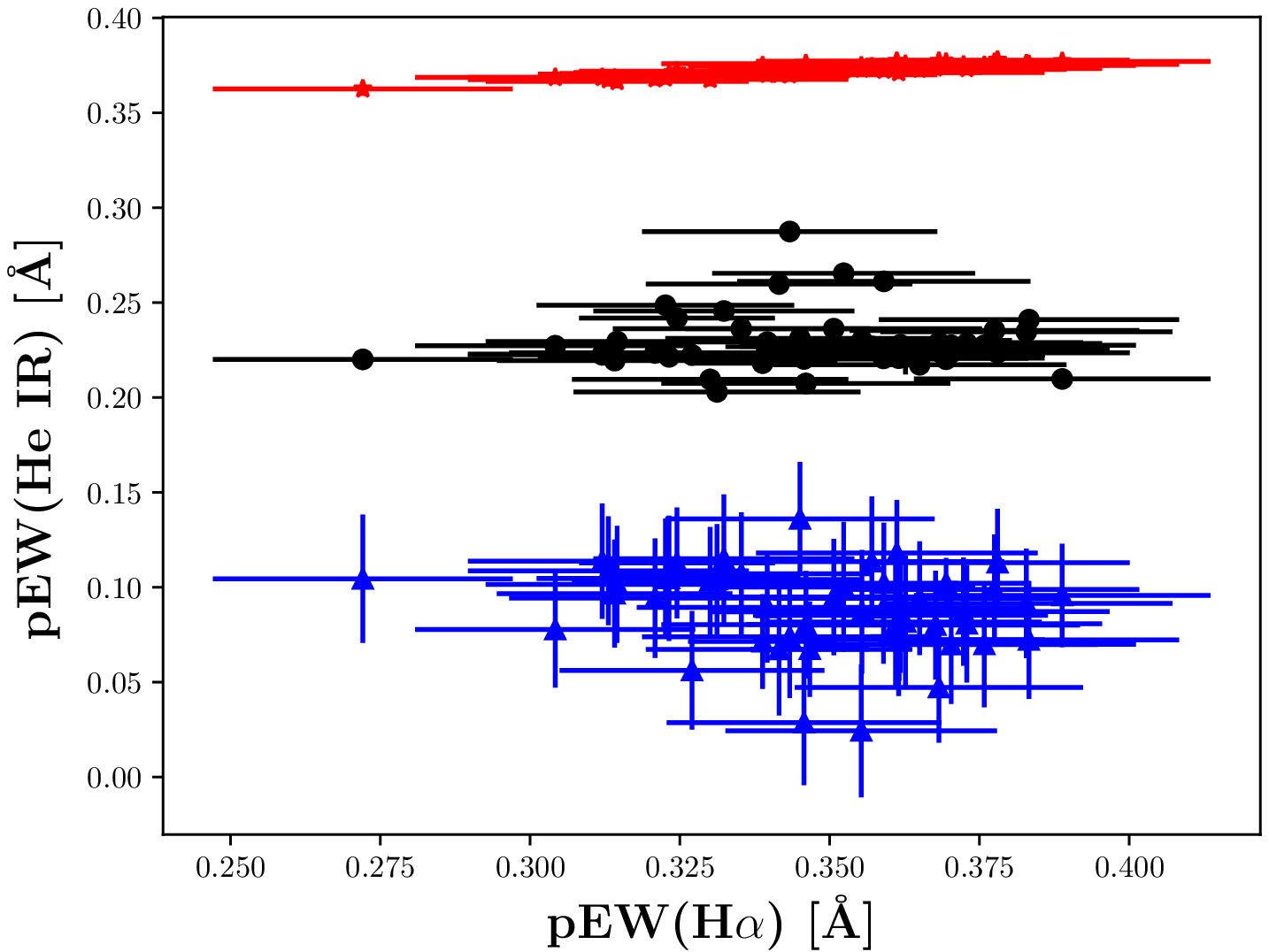}\\
\caption{\label{novar} All available spectra of the M1.5\,V star GJ~2. \emph{Top:}
  The region around the \heir\ line. Some artefacts of
  airglow and cosmics manifest themselves as emission spikes. \emph{Middle:} The region around
  H$\alpha$. Minor telluric contamination at about 6559-6560 and 6566-6567 \AA\ is evident.
  The dashed vertical lines indicate the central wavelength of the \heir\ and H$\alpha$ lines,
  respectively. \emph{Bottom:} Correlation for different chromospheric lines to pEW(H$\alpha$). Black denotes
  pEW(\heir), blue denotes pEW(\hei\ D$_{3}$) scaled by -4, and red denotes pEW(\cab) scaled by 0.5. }

\end{center}
\end{figure}

To investigate the uncertainty in the pEW measurements and the resulting
MAD values, we looked at the lowly active M1.5\,V star J00051+457 (GJ~2).
Since GJ~2 shows a typical number, strength, and position of artefacts, we used it
as a reference.
The H$\alpha$ line spectra of GJ~2 display variability at a low level
(Fig.~\ref{novar}) and
there is a correlation between the pEW(H$\alpha$) and pEW(\cab)
with a value of 0.69 for Pearson's correlation coefficient (\mbox{p$=
5.5\,10^{-6}$}).
However, there is no correlation with the pEWs of the \heir\ line because
it is not sensitive enough, as we discuss below. We deem the \hei\ IR
line of GJ~2
intrinsically stable.

The statistical uncertainty of pEW(\heir) is  small as can be seen in
Fig.~\ref{novar}, where the error bars are smaller than the used symbols.
Indeed the
relative error is about one percent; while for pEW(H$\alpha$) it is about six percent.
If this were the only source of variation, this should
be reflected by the standard
deviation of the values in the time series. However,
the latter value is about four times higher than the statistical
uncertainty, demonstrating
that the variation of the pEW measurements is dominated by other effects
caused by, for example telluric correction, cosmics, instrumental effects, or
normalisation problems.

To obtain
an error estimate on the MAD, we used the bootstrap method. In particular,
we generated bootstrap samples
by randomly selecting pEW measurements (with replacement) from the time
series and computed the associated
MAD. From the thus obtained distribution, we computed the standard deviation
of the MAD, which amounts to
about 20 \% of its original value as well for MAD(\heir) as for
MAD(H$\alpha$). We consider this a lower limit for the achievable precision of the MAD
estimate.
For stars with fewer observations, the uncertainty on the MAD scales
approximately inversely proportional
to the square root of the number of data points, reaching around
$0.5\times$MAD for stars
with less than ten spectra. Although the details depend on the individual
case,
we conclude that an uncertainty estimate of $0.2\times$MAD is
applicable to the majority of our stars.
For the latest sub-type stars with high levels of statistical noise in their
spectra and for some
stars with much stronger artefacts this estimate may be too low.

\begin{table}
\caption{\label{ew} Parameters of the pEW calculation. }
\footnotesize
\begin{tabular}[h!]{lcccc}
\hline
\hline
\noalign{\smallskip}

               & Line   & full  & reference band 1 & reference band 2 \\
               &        & width & \\
& [\AA] & [\AA] & [\AA] & [\AA]\\
\noalign{\smallskip}
\hline
\noalign{\smallskip}
\heir& 10833.26 & 1.6 & 10820.0--10822.5 & 10851.0--10853.0 \\
H$\alpha$& 6564.62 & 1.6 & 6537.43--6547.92 & 6577.88--6586.37 \\
\cab  & 8500.33 & 0.5 & 8476.33--8486.33 & 8552.35--8554.35\\
\ion{He}{i}\,D$_{3}$ & 5877.24 & 1.0 & 5870.00--5874.00& 5910.0--5914.0\\
\noalign{\smallskip}
\hline

\end{tabular}
\normalsize
\end{table}

\begin{sidewaystable}
\caption{\label{measurements} Measured mean pEWs, MADs, and correlation coefficients of the considered lines.$^{a }$}
\footnotesize
\begin{tabular}[h!]{lcccccccccccccccc}
\hline
\hline
\noalign{\smallskip}
Karmn       &corr & p-val& corr &p-val& corr  & p-val& corr & p-val& mean & mean& mean& mean & MAD  &MAD & MAD & MAD\\
            &     \heir-            &      & \heir-         &     &   \heir-  &    &  H$\alpha$- &   &   pEW & pEW& pEW&pEW &(\heir)&(H$\alpha$)&(\hei D$_{3}$)&(\cab)\\        
            &      H$\alpha$        &      & \hei D$_{3}$    &     &    \cab  &    &    \cab     &    &  (\heir) & (H$\alpha$)   &  (\hei D$_{3}$) & (\cab) &  &  &  & \\ 
            &                       &      &                 &     &          &    &             &    &  [\AA] & [\AA] & [\AA]& [\AA]  & [\AA] & [\AA] & [\AA] & [\AA]\\      
\hline
J00051+457  &    0.215 &  0.148 &  0.291 &  0.047 &  -0.030 & 0.843 &  0.609 &  0.000 &  0.225 &  0.343 &  -0.026 &  0.748 &  0.004 &  0.019 &  0.003 &  0.006\\
J00067-075  &    -0.029&  0.813 &  -0.111&  0.362 &  0.290  & 0.015 &  0.427 &  0.000 &  0.017 &  -0.071&  0.053  & 0.402  & 0.004  & 0.050  & 0.024  & 0.009\\
J00162+198E &    0.269 &  0.484 &  -0.350&  0.356 &  0.342  & 0.368 &  0.878 &  0.002 &  0.065 &  0.094 &  -0.016 & 0.553  & 0.014  & 0.011  & 0.010  & 0.007\\
J00183+440  &    0.443 &  0.000 &  -0.217&  0.003 &  0.037  & 0.615 &  0.136 &  0.063 &  0.105 &  0.311 &  -0.003 & 0.793  & 0.005  & 0.007  & 0.004  & 0.002\\
J00184+440  &    0.560 &  0.000 &  -0.583&  0.000 &  0.862  & 0.000 &  0.651 &  0.000 &  0.025 &  0.154 &  0.049  & 0.765  & 0.005  & 0.014  & 0.005  & 0.002\\
J00286-066  &    -0.673&  0.002 &  0.095 &  0.708 &  -0.595 & 0.009 &  0.568 &  0.014 &  0.075 &  0.134 &  -0.021 & 0.536  & 0.005  & 0.015  & 0.011  & 0.004\\
J00389+306  &    0.149 &  0.568 &  0.006 &  0.982 &  -0.345 & 0.175 &  0.387 &  0.125 &  0.139 &  0.281 &  -0.026 & 0.743  & 0.006  & 0.015  & 0.007  & 0.007\\
J00570+450  &    -0.309&  0.456 &  -0.162&  0.702 &  -0.575 & 0.136 &  0.752 &  0.031 &  0.118 &  0.130 &  -0.028 & 0.671  & 0.005  & 0.034  & 0.005  & 0.004\\
J01013+613  &    -0.071&  0.856 &  -0.232&  0.548 &  -0.151 & 0.698 &  0.983 &  0.000 &  0.133 &  0.203 &  -0.026 & 0.737  & 0.005  & 0.047  & 0.003  & 0.003\\
J01019+541  &    0.262 &  0.294 &  0.295 &  0.235 &  0.301  & 0.225 &  0.971 &  0.000 &  0.136 &  -4.462&  -0.813 & 0.183  & 0.038  & 0.447  & 0.072  & 0.024\\
\multicolumn{17}{c}{\ldots}\\ \hline
\end{tabular}

$^{a}$ The full table is provided at CDS. We show the first 10 rows as a guideline. 
\normalsize
\end{sidewaystable}

%
%
%

\section{General \heir\ line variability properties}\label{sec:generalvar}

In this section we briefly summarise our findings about the time-average
properties of the \heir\ line from \citet{hepaper}, which represent the quiescent state of the star.
In the latter paper we find that the observed strength of the line
depends on the spectral type or effective temperature: In early-type M dwarfs it manifests itself ubiquitously as a strong 
absorption line with pEW(\heir) up to 300 m$\AA$, which declines to later spectral sub-types and
becomes undetectable with the methods used by \citet{hepaper} at about M5
(and pEW(\heir) < 50 m$\AA$ for the stars with the best signal-to-noise ratio).
We attribute the few cases
in which the line is found in emission in the average spectrum to flaring activity strong or
frequent enough to affect the average spectrum.
 
\subsection{\heir\ line variability}\label{sec:linevar}

\begin{figure}
\begin{center}
\includegraphics[width=0.5\textwidth, clip]{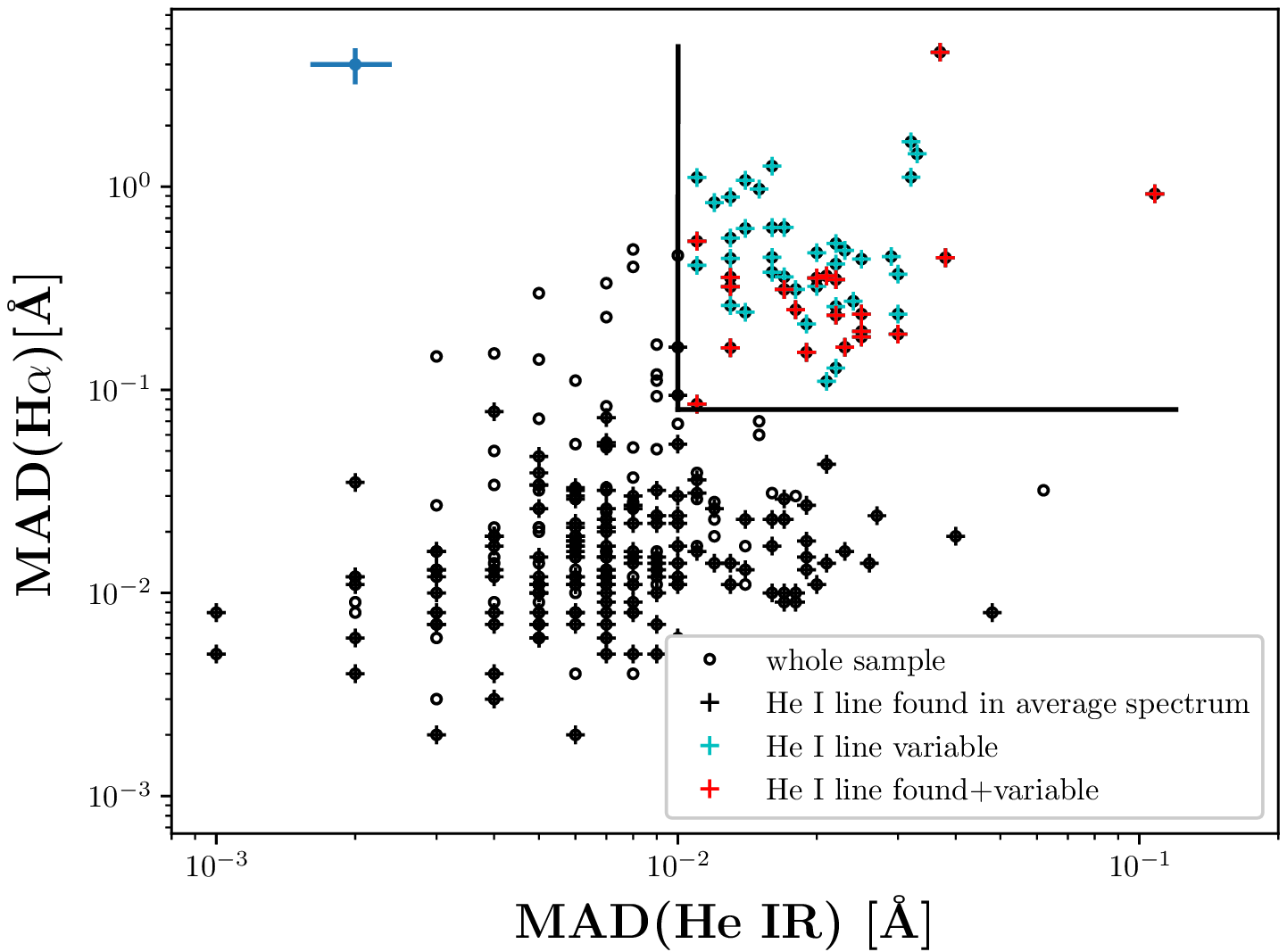}\\
\caption{\label{madvsmad} Median average deviation (H$\alpha$) as a function of MAD(\heir) for the sample stars (black open circles).
  The black crosses indicate stars for which the \heir\ line was detected in the averaged spectra by \citet{hepaper}.
  The cyan crosses indicate stars where the line could not be found in the average spectrum
  but is detectable by \heir\ line variability,
  and the red crosses denote stars with both properties. The error bars shown at the top left corner illustrate
  the 20 \%\, systematic error estimated as described in Sect. \ref{sec:equivwidth}.
  The black vertical and horizontal lines denote the thresholds
  for our conservative variable \hei IR line sample as defined in Sect. \ref{conservative}.
}
\end{center}
\end{figure}

For an automatic detection of variability we looked at the scatter in the data that we measured
with the MAD.  In Fig.~\ref{madvsmad} we compare the MADs of the pEW measurements of the H$\alpha$ and \heir\ lines.
The star with the largest MAD(H$\alpha$) is J05084-210 (2M~J05082729-2101444) and the star with the highest
MAD(\heir) is J01352-072 (Barta~161~12); both are young stars (see Table \ref{variablestars})
 and are discussed in Sect. \ref{examples} in more detail. 
  As seen in Fig.~\ref{madvsmad},
the pEW scatter in the H$\alpha$ line is about 30 times larger than
that in the \heir\ line, while the integration band widths of $1.6$~\AA\ are identical (Table~\ref{ew}).
Therefore, the H$\alpha$ line is considerably more variable in each time series and, as a consequence, it also shows larger
variations within our sample of stars and is thus also a more sensitive activity
indicator than the \heir\ line.
Figure \ref{madvsmad} shows that most stars for which the
  \heir\ line could be detected in the average spectra by \cite{hepaper} do not show
  variability, while there are a number of stars for which the line could not be detected
  in the average spectra, but reveals itself now by variability. This latter case is
discussed below.

Moreover, Fig. \ref{madvsmad} shows that there are two groups of stars: an inactive group at the bottom
left and an active group with higher variability at the top right. 
The majority of our sample stars belong to the inactive group and show only
low levels of variability in the pEW(H$\alpha$) as measured by MAD(H$\alpha$) < 0.1 \AA.
These sample stars nevertheless may display the full range of observed MAD(\heir).
In particular, there are five stars with MAD(\heir) greater than 0.038~\AA\, and low MAD(H$\alpha$).
Visual inspection shows that many spectra of these stars are
subject to artefacts of cosmic rays and airglow
in the spectral range of the \heir\ line, thereby leading to a large spurious MAD(\heir) without physical reason but caused
by long exposure times because all five are rather faint stars.
In contrast, all 11 stars with MAD(H$\alpha$) > 0.75~\AA\,
are also found variable in the \heir\ line. Before we define our variability criterion in the next section,
we want first to present an example of an inactive and an active star.

To give an example for line variability in inactive stars, we refer again to Fig. \ref{novar}
  in which we show all available \heir\
and H$\alpha$ line spectra of 
the weakly active star GJ~2. The correlation behaviour of the lines
was already discussed in Sect. \ref{sec:equivwidth}.  
The spectra show a prominent \heir\ line along with 
a \ion{Si}{i} line at 10\,830.057~\AA\, and
a \ion{Na}{i} line at 10\,837.814~\AA.
All other spectral features
in the shown wavelength range remain unidentified; see a discussion in \citet{hepaper} and compare to
\citet{Andreasen} and \citet{Marfil2020}.
Figure~\ref{novar} also shows that telluric lines are a minor problem around the H$\alpha$ line. 
As can be seen from Table \ref{measurements}, the MAD(H$\alpha$) and MAD(\hei) are
0.019 and 0.004~\AA, respectively, and therefore well within the cloud at the lower left
seen in Fig. \ref{madvsmad}, which should represent stars with low levels of activity.



\begin{figure}[h!]
\begin{center}
\includegraphics[width=0.5\textwidth, clip]{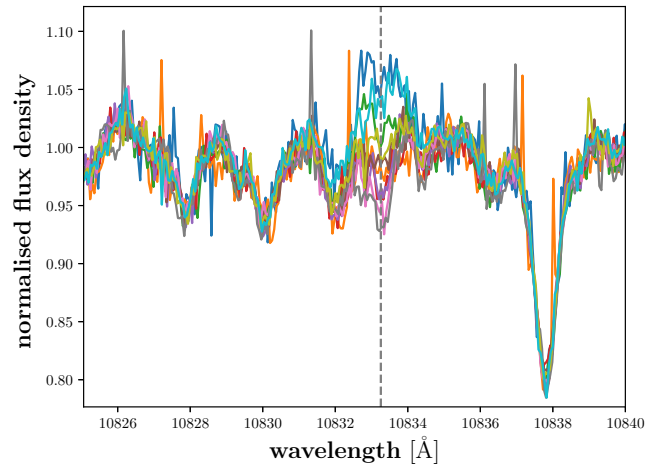}\\
\includegraphics[width=0.5\textwidth, clip]{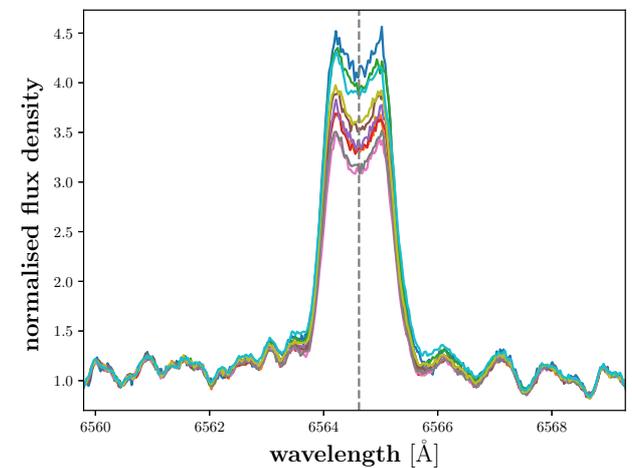}\\
\includegraphics[width=0.5\textwidth, clip]{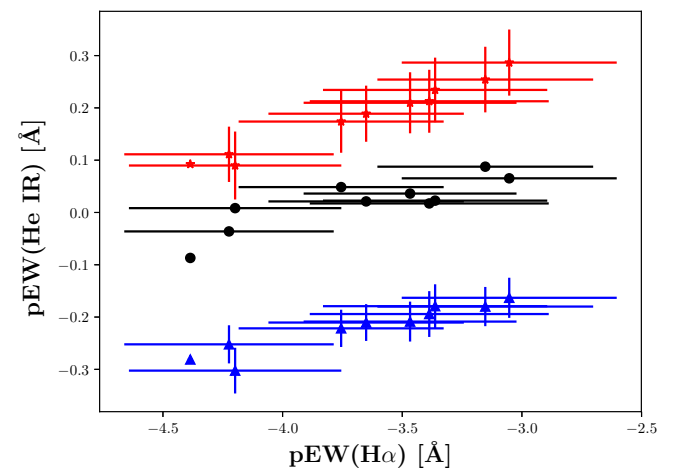}\\
\caption{\label{J03473} All available spectra of the star G~080-021. \emph{Top:}
  The region around the \hei\ IR line. \emph{Middle:} The region around
  H$\alpha$. The dashed vertical lines indicate the central wavelength for the \heir\
  and the H$\alpha$ lines, respectively. \emph{Bottom:} Correlations for different chromospheric lines to pEW(H$\alpha$);
 black denote pEW(\heir), red pEW(\hei\ D$_{3}$) scaled by 0.5,  and blue pEW(\cab).
}
\end{center}
\end{figure}

In contrast the M3.0 star J03473-019 (G~080-021) provides
an example of a more active star. The nine available CARMENES spectra, as can be seen in Fig.~\ref{J03473},
show that the H$\alpha$ line is always in emission with obvious variability. The \heir\ line also
displays clear variability going from absorption into emission.  As a consequence, the
line is almost absent in the average spectrum, which is consistent with our previous
non-detection in \citet{hepaper}.
The pEW(\heir) is correlated with all of pEW(H$\alpha$); pEW(\hei D$_{3}$),which is in emission; and pEW(\cab). Also, pEW(H$\alpha$) and pEW(\cab) are correlated.
All correlation coefficients are higher than 0.94 with p \mbox{$ < 10^{-4}$}.
Concerning the variability as measured by the MAD(H$\alpha$) and MAD(\hei) of
0.273 and 0.024~\AA, respectively, the star is located in the top part of Fig.~\ref{madvsmad}.

\subsection{Conservative \hei\ variable sample}\label{conservative}

Visual inspection proves that for some stars the identification of
variability in the \hei\ IR line
remains ambiguous, mostly because of the artefacts found in the line region.
Therefore, we proceed to
identify the stars showing the most reliably physical variation in the \heir\ line. 
To this end, we
demand that both the H$\alpha$ and \heir\ lines show significant scatter in their pEW measurements. 
We define the conservative \hei\ variability sample (conservative sample for short)
to comprise all stars with MAD(H$\alpha$)$ > 0.08$ and MAD(\heir) $> 0.01$\AA.
This roughly corresponds to the borderlines between the clouds of the least active stars at the bottom left
and the most active stars at the top right in Fig.~\ref{madvsmad}.
A total of 56 stars (18\,\% of the sample) show variability on this level.
This conservative sample is very similar to the
stars we found by visual inspection of their spectra to be variable, therefore emphasising the robustness
of the method. Moreover, we used the visual inspection to fine-tune the thresholds in MAD(H$\alpha$) and MAD(\heir).

All stars in the conservative sample show
H$\alpha$ in emission, which we define by pEW(H$\alpha$) $< -0.6$~\AA\ following \citet{hepaper}.
In fact, only such stars match the MAD(H$\alpha$) > 0.08 criterion.
About 80\,\% of the H$\alpha$ emitters are included in the conservative sample
as well as approximately the same fraction of the 
fast rotators (\vsini $> 25$~km\,s$^{-1}$). On the other hand,
for most stars in which we find the \heir\ line to be variable, this was not found 
in the averaged spectra in our previous study and
the mean pEW(\hei) differs only marginally from zero.
This indicates that either the line varies from absorption to emission in these stars and, therefore,
is hardly visible in the average spectrum, or the line is not present in the spectra during quiescence and only evolves
into emission during flares. For the first case, we already showed an example in Fig.~\ref{J03473},
while for the second case more examples are presented in Sect.\ref{examples}.

\begin{figure}
\begin{center}
\includegraphics[width=0.5\textwidth, clip]{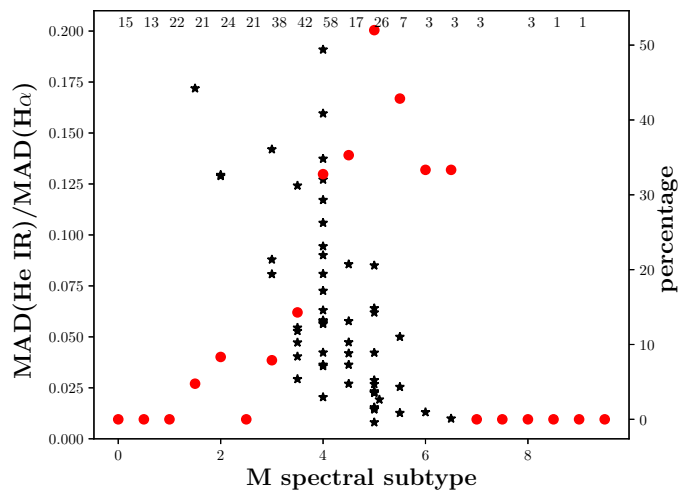}\\
\caption{\label{percentage} Median average deviation (\hei)/MAD(H$\alpha$) as a function of spectral sub-type (black asterisks).
  The dots for the two M2.0 stars lie on top of each other. The red dots
  correspond to the
  right axis where we give the percentage of stars found to be variable. The total number of 
  stars for each bin are given at the top of the panel.
}
\end{center}
\end{figure}

The earliest star in the conservative sample is the M1.5 dwarf
J15218+209 (OT~Ser), which is known to be young \citep{Shkolnik09}  and active \citep{Shulyak2019},
followed by the two M2.0 dwarfs J09425+700 (GJ~360) and J11201-104 (LP~733-099).
These two latter stars are significantly variable in H$\alpha$,
but visual inspection shows that the \heir\ line is again subject to artefacts.
This lack of \heir\ line variability in early M dwarfs is connected to the general inactivity
of these stars. 
Although they display a chromosphere, its contribution to the spectrum
is not strong and variable enough (or both) to be significant in this type of analysis.
At spectral type M3.0, a steep increase in the number and fraction of variable stars occurs, which reaches a peak at M4.0 
and then decreases again towards M6--6.5. We show this behaviour in Fig. \ref{percentage}, together
with the behaviour of the ratio of MAD(\hei) and MAD(H$\alpha$).
Overall, the pEW variation as measured by the MAD is about an order of magnitude larger in H$\alpha$ than
in the \hei\ IR line. The distribution appears to show a trend to
lower ratios for later-type
stars indicating that the variability in the \hei\ IR line declines relative to that in H$\alpha$.
This may correspond to the decline of the line strength in \hei\ IR found in our previous study.
Nevertheless, we find that out 
of the 23 M5.0 dwarfs, 13 match the criterion of the conservative sample,
although \citet{hepaper} was able to find the \heir\ line in the average stellar spectra for only one of these stars.
Among the
22 stars with later spectral types, we found variations in only two owing to the increasing noise in the
spectra.


\subsection{Correlation between pEW values for different chromospheric lines}
For most of the 56 stars in the conservative sample,
pEW(H$\alpha$) is well correlated to pEW(\cab)
with a mean Pearson's correlation coefficient of $r$=0.87.
Furthermore, 30 stars also show a correlation between pEW(H$\alpha$) and pEW(\hei) with $r$ > 0.6 and a
p-value below 0.05 (which we regard as the highest p-value for a significant correlation). Although
many of these stars show prominent signs of flaring in at least some of their spectra,
there are also a few spectral time series without clear flare signatures (see Fig.~\ref{J03473}).

\begin{figure}
\begin{center}
\includegraphics[width=0.5\textwidth, clip]{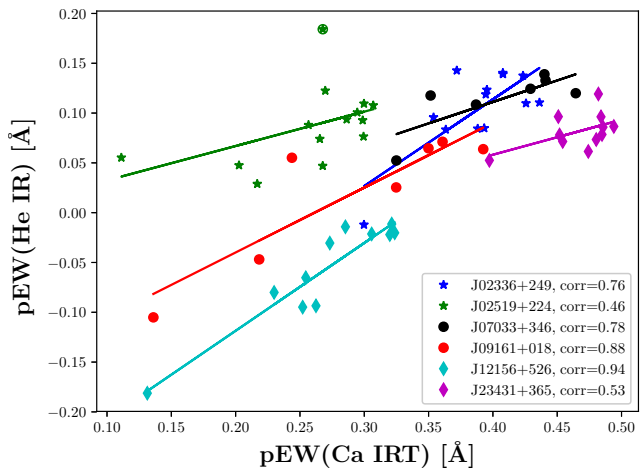}\\
\caption{\label{correlation} Pseudo-equivalent width(\hei) as a function of pEW(\cab) for six representative
  M4.0 stars. Measurements are designated by the symbols, while linear fits to the measurements are
  given by the solid lines. The CARMENES identification numbers of the stars are given in the legend together
  with the respective Pearson's correlation coefficient. The outlier caused by a flare for J02519+224/RBS~365
  is denoted with an additional circle.
}
\end{center}
\end{figure}

To illustrate the correlations better we show the time series of six stars out of the
19 M4.0 dwarfs in Fig. \ref{correlation}. These  stars were selected for their representative number
of spectra and a broad coverage in pEW(\cab) and pEW(\hei). Besides a few stars with low or negative
correlation values that have non-significant p-values, all the M4.0 dwarfs show similar slopes of their
linear approximations. Many stars have one or two outliers in their pEW(\hei). For example, for J02519+224 (RBS~365)
the Pearson's correlation coefficient is 0.46, and 
amounts to 0.63 without the outlier in pEW(\hei), which is caused by a flare.
The similarity in the slopes is not necessarily expected, since our variability criterion is not
tied to correlation but only to variation. Since H$\alpha$ is thought for M dwarfs to first deepen
with increasing activity levels before going into emission \citep{Cram1979}, it is possible to also naively
expect negative correlation. For
the Sun and solar-like stars the correlation coefficient
between H$\alpha$ and the \ion{Ca}{ii}~H and K lines for example may vary from -1 to 1 for different stars.
This was explained
by \citet{Meunier} mainly by different filling factors and contrasts between plages and filaments. 
In our sample no significant negative correlation was found. Moreover, all 15 stars among the
conservative sample with $r$ < 0.5 for the correlation
between pEW(H$\alpha$) and pEW(\hei) turned out to be caused by outliers either because of
low signal-to-noise ratios, airglow artefacts, or even flaring (in case the lines do not react
in the same fashion to the flare).
Discarding these outliers leads to significant correlation of the time series. Therefore, we conclude,
that variability in the \heir\ line always comes with a positive correlation to pEW(H$\alpha$).

\subsection{Rotational variability}

Besides the strong variability detected by our method 
there may be much lower amplitude variations in the sample that
are nevertheless correlated, if they are caused for example by the plage rotating with the star.
Rotational modulation is a source of variability in chromospheric indicators and provides one
possibility to measure rotation periods  \citep[][]{Mittag2017, rotation}.
Unfortunately, we have fewer than 30 spectra for most of
the stars showing a high correlation between pEW(H$\alpha$) and pEW(\hei),
so that a period search is not promising. The low number of spectra for these stars is due to
the exoplanet search goal of the survey: active stars were observed a few times and then
discarded from the survey after finding that their activity causes radial velocity jitter
\citep{Lev}.

There are
20 stars showing values larger than $0.7$ for Pearson's correlation
coefficient between pEW(H$\alpha$) and pEW(\heir) (irrespective of the p-value),
which are not included in the conservative sample; these stars may exhibit a low level variation
that is not detected by our method.
However, the majority of these stars suffer again from artefacts or a low number of spectra.
We only consider the variation real for three stars:
the M5.0\,V star J06318+414 (LP~205-044),
the M1.0\,V star J10251-102 (BD-09~3070), and the M3.0\,V star
J19084+322 (G~207-019); the latter two have H$\alpha$ in absorption.
However, LP~205-044, which is discussed later among the
exceptional examples in Sect.
\ref{examples}, shows H$\alpha$ in emission and is a fast rotator.
However, if the low level variation of these three stars is caused by rotational
modulation stays elusive.


For a further search for rotational modulation
we also  revisit the cases of
J13536+776 (NLTT~35712) and J01026+623 (BD+61~195) to follow up previous reports on rotational
modulation in their chromospheric lines.
For the M4.0\,V star NLTT~35712, \citet{Dupree2018}
report \heir\ line variation that is consistent with
rotational modulation with the known period of 1.231~d \citep{Newton2016}.
The star is included in our conservative sample. It shows a high value of 0.77 for
Pearson's correlation coefficient between pEW(\heir) and pEW(H$\alpha$), but a rather large p-value of 0.08.
Looking at the \heir\ line spectra reveals that one spectrum shows an artefact and another spectrum was taken during a flare
as evidenced by a blue
asymmetry in H$\alpha$ \citep{asym}. The seven available spectra were taken between April 2016 and January 2018
each at least about a month apart. In a phase folded time series pEW(H$\alpha$) and pEW(\hei) do not show a conclusive
pattern. This inconclusiveness together
with the high amplitude of changes in H$\alpha$ outside flares seems to
indicate that a possible variation with the rotation period is at least veiled if not dominated by intrinsic
variability in these lines on the timescales covered in this study.

As a further check for the sensitivity of the \heir\ line to rotational modulation, we examine the M1.5\,V
star BD+61~195, whose period is known to be 18.4~d \citep[][obtained from spectroscopic lines]{SM2017}
to 19.9 d \citep[][obtained photometrically]{DA19}. This star also represents the most benevolent case
in our own previous period search based on CARMENES activity indices \citep{rotation}, where   
we recovered the rotation period in the pEW(H$\alpha$) and pEW(\cab) using a generalised Lomb-Scargle periodogram
as implemented in \texttt{PyAstronomy} \citep{Zechmeister2009, Czesla2019pya} with relative ease.
In contrast, no significant period can be established for the pEW(\heir) time series.
We consider this a likely consequence of the inferior sensitivity of the \heir\ line compared to
the other indicators.

\section{Individual examples of variability}\label{examples}
\subsection{Classification of the examples}

\begin{table*}
\caption{\label{variablestars} Stellar parameters of the discussed variable stars }
\footnotesize
\begin{tabular}[h!]{llcccll}
\hline
\hline
\noalign{\smallskip}
Karmn       &name & spectral & \vsini & P$_{\rm rot}$& moving  \\
            &     &  type    &   [km\,s$^{-1}$]   & [d]       &  group \\
\hline
\noalign{\smallskip}
J01352-072 & Barta~161~12    & M4.0 V (Ria06) & 59.8 $\pm$ 6.9 (Rein18) & 0.7 (Kira12) & $\beta$\,Pic (Malo14)&  ...\\
J02519+224 & RBS~365         & M4.0 V (Ria06) & 27.2 $\pm$ 2.7 (Rein18) & 0.86 (DA19)   & LA (CC20)  \\
J05084-210 & 2MASS~J05082729-2101444 & M5.0 V (Ria06) & 25.2 $\pm$ 2.5 (Rein18) & ...    & $\beta$\,Pic (Malo14)\\
J06318+414 & LP~205-044      & M5.0 V (PMSU) & 58.4 $\pm$ 26.1 (Rein18) & 0.30 (DA19) & LA (CC20)\\
J07472+503 & 2MASS~J07471385+5020386 & M4.0 V (Lep13) & 10.1 $\pm$ 1.5 (Rein18) & 1.32$\pm$ 0.01 (DA19) & UMa (CC20)\\
J11474+667 & 1RXS~J114728.8+664405   & M5.0 V (AF15) & 2.7 $\pm$ 1.5 (Rein18) & ... &  Cas (CC20) \\
J22231-176 & LP 820-012      & M4.5 V (PMSU)  & < 2 (Rein18) & ... & ...  \\
J22468+443 & EV~Lac          & M3.5 V (PMSU)  & 5.9 $\pm$ 0.1  (Fou18) & 4.38 $\pm$ 0.03 (DA19) & UMa (CC20) \\
J22518+317 & GT~Peg          & M3.0 V (PMSU)  & 13.2 $\pm$ 0.9 (Fou18) & 1.63 $\pm$ 0.01 (DA19) & ...\\

\noalign{\smallskip}
\hline

\end{tabular}\\
\tablebib{
  AF15:~\citet{AF15a}; CC20:~\citet{CC19}; DA19:~\citet{DA19}; Fou18:~\citet{Fou18};   Kira12:~\citet{Kiraga2012};
  Lep13:~\citet{Lepine};  Malo14:~\citet{Malo2014}; 
  PMSU:~\citet{PMSU}; Rein18:~\citet{Reiners2017}; Ria06:~\citet{Ria2006}
  
}\\

\normalsize
\end{table*}

\begin{table}
\caption{\label{dateofflares} Julian date of major flares for the stars from Table \ref{variablestars} }
\footnotesize
\begin{tabular}[h!]{llll}
\hline
\hline
\noalign{\smallskip}
Karmn       & JD of flare & colour  & reference\\
            & -2450000 [d] & in Fig. &number\\
\hline
\noalign{\smallskip}

RBS~365    & 7693.50903 & blue & 1\\
           & 7985.61434 & orange & 2\\  
LP~205-044 & 7823.44073 & blue & 1\\
           & 8057.63874 & orange & 2\\
J07472+503 & 8005.66491 & blue& 1\\
           & 8031.64099 & orange & 2\\
           & 8081.56145 & cyan & 3\\
J11474+667 & 7762.54635 & blue & 1\\
EV Lac     & 7632.62893 & blue & 1\\
           & 7633.46711 & orange & 2\\
           & 7650.53689 & green & 3\\
           & 7968.42107 & red & 4\\
           & 7999.35888 & purple & 5\\
           & 8032.42971 & brown & 6 \\
GT Peg     & 7762.2768 & blue & 1\\
\noalign{\smallskip}
\hline

\end{tabular}\\
\normalsize
\end{table}

Since not much is known about the behaviour of the \heir\ line in M~dwarfs during flaring activity, in this section we
present some examples of exceptional line profiles. These
examples were selected by visual inspection and the list is not complete with respect to any definitive criterion.
Nevertheless, we tried
to cover a broad variety of different line profiles, which seem to be typical reactions to different flare phases.
Since we only have snapshots, flare identification is not always certain.
Therefore, we restricted our examples to spectra for which we identify flaring activity by line
broadening in H$\alpha$.
Flares with either symmetric or asymmetric broadening in the H$\alpha$ line almost always show an extraordinary
reaction in the \heir\ line while flares that manifest themselves only in an amplitude enhancement
in H$\alpha$ typically show minor reactions in the \heir\ line.

In the case of asymmetric H$\alpha$ line broadening, the asymmetry helps to determine the flare phase.
In particular,
blue asymmetries are thought to be caused by chromospheric evaporation
associated with flare onset and red asymmetries are thought to be caused by coronal rain occurring
during the decay phase \citep{asym}. Symmetric broadening may be attributable to 
Stark broadening, turbulence, or an integration effect. Our typical exposure times of 15~min are  long
compared to the flare timescale so that both signatures of chromospheric
evaporation and back-falling material may be caught in the same spectrum.
Turbulent broadening may also be responsible for the broadening
in asymmetric cases, when the moving material is additionally turbulent.

We sort our examples by the type of broadening observed in H$\alpha$ and thereafter
present two exceptional stars, J01352-072 (Barta~161~12)  and J06318+414 (LP~205-044), which are most puzzling.
In Table~\ref{variablestars},
we give the spectral type, \vsini, the rotation period,
and the moving group the stars belong to (if applicable)  for all stars for which we show spectra in this section. 
For better reference we identify all major flares shown by the Julian date of their occurrence in Table 
\ref{dateofflares}.

\subsection{Blue asymmetries in H$\alpha$}

\begin{figure*}
\begin{center}
\includegraphics[width=0.5\textwidth, clip]{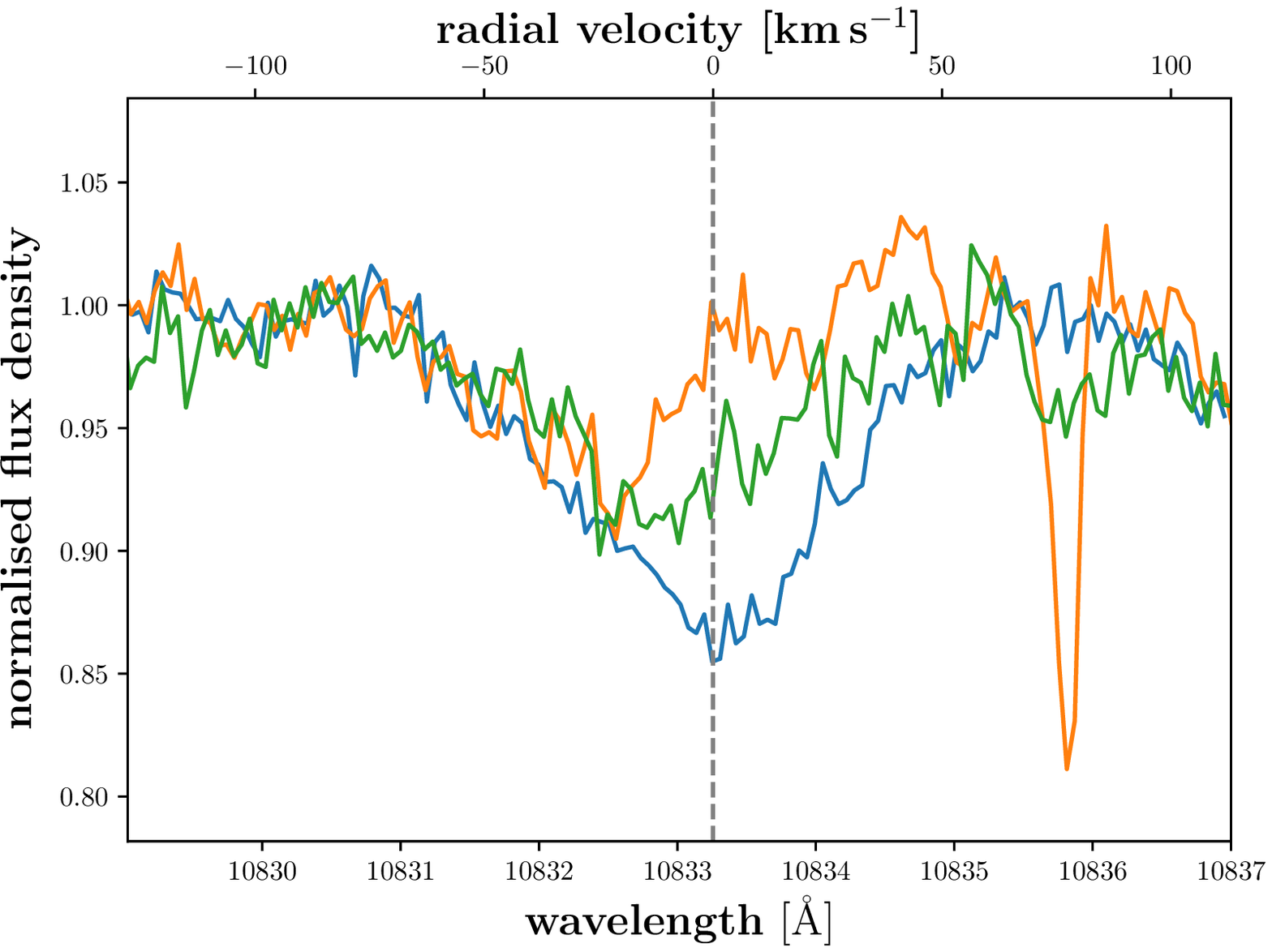}
\includegraphics[width=0.5\textwidth, clip]{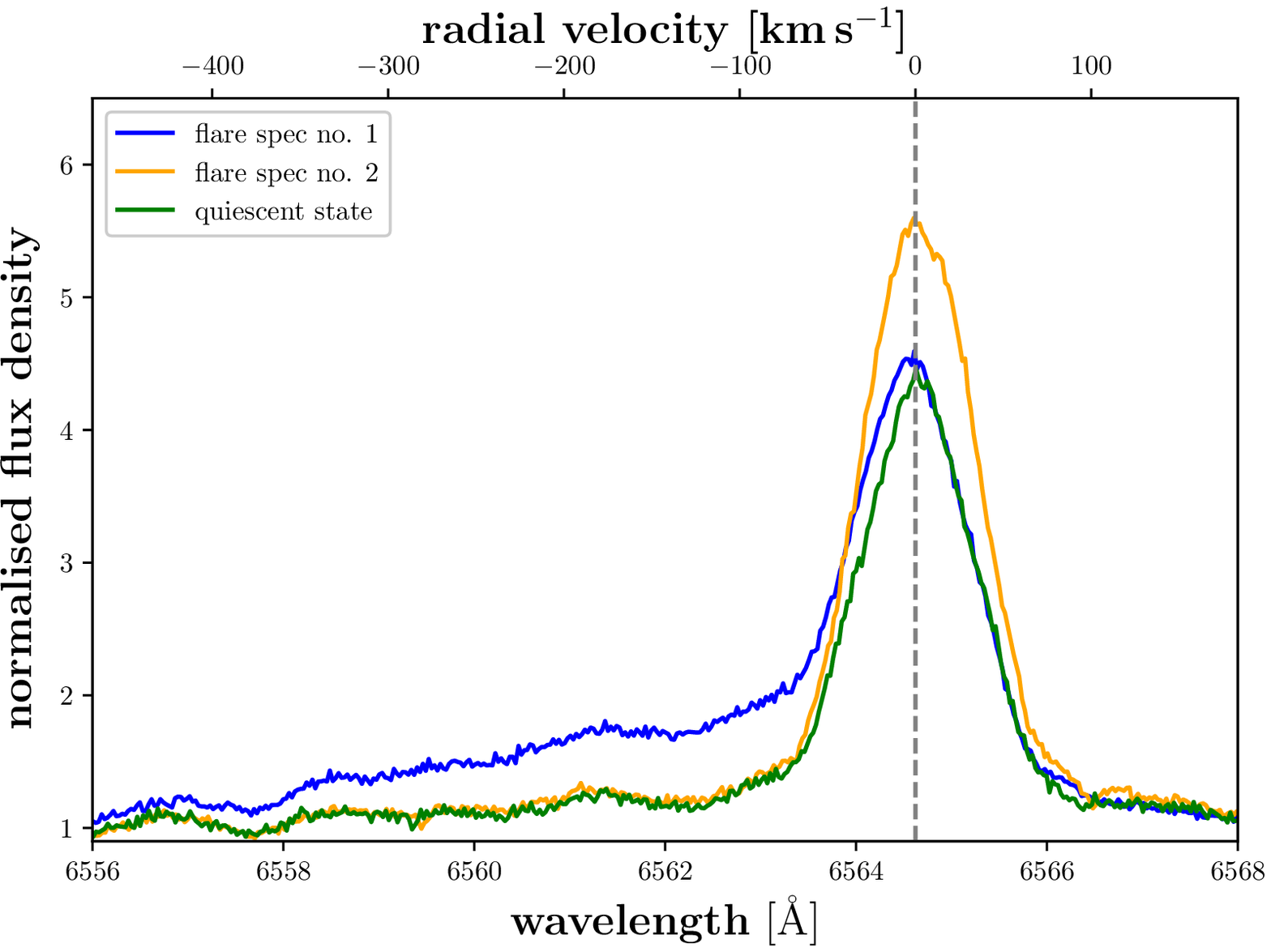}\\
\caption{\label{variabilityRBS365} Three of 14 spectra for different chromospheric indicator
  lines for  RBS~365. \emph{Left:} The \hei\ IR line. \emph{Right:} The H$\alpha$
  line. The spectrum with the pronounced blue wing in H$\alpha$ corresponds to the deepest
  spectrum in \hei \ IR (both indicated by the blue solid line corresponding to flare spectrum no. 1).  
  The green spectra represent the most quiescent
  case and the orange line a more active state (flare spectrum no. 2), which may or may not be associated with flaring.
  The dashed vertical lines denote
  the central wavelength of the displayed chromospheric lines.
}
\end{center}
\end{figure*}

\begin{figure*}
\begin{center}
\includegraphics[width=0.5\textwidth, clip]{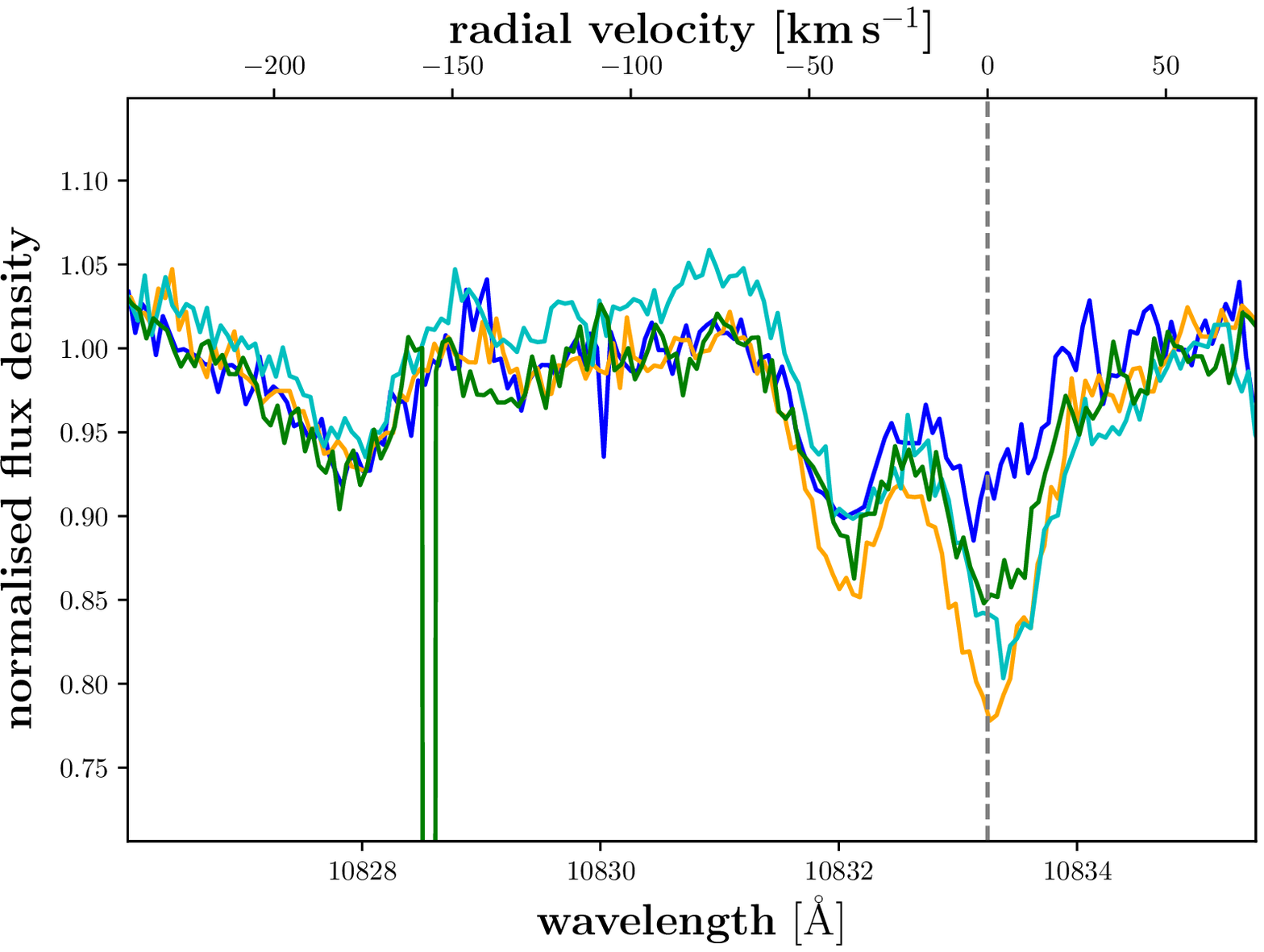}
\includegraphics[width=0.5\textwidth, clip]{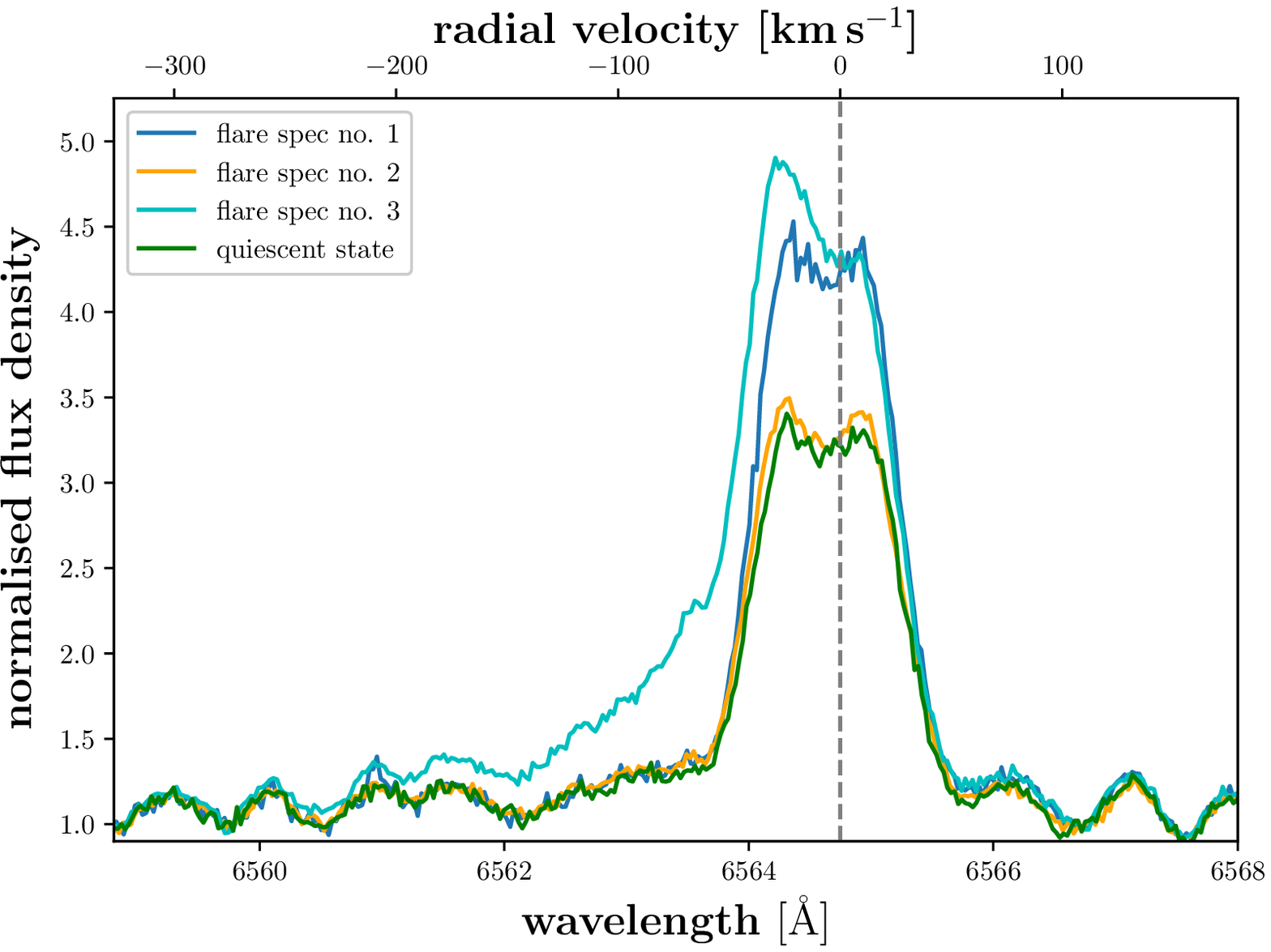}\\
\caption{\label{variabilityJ07472} Four out of six available spectra for different chromospheric indicator
  lines for 
  J07472+503. \emph{Left:} The \hei\ IR line. \emph{Right:} The H$\alpha$
  line. 
The dashed vertical lines indicate
the central wavelength of the displayed chromospheric lines. The same colours indicated spectra corresponding
to the same observation date; the corresponding dates are given in Table \ref{dateofflares}. The emission wing of the blue spectrum is most clearly seen at around
10\,831 \AA. The green spectrum represents the most inactive state.}
\end{center}
\end{figure*}

As examples of stars with prominent H$\alpha$ blue asymmetries, we show the spectra of J02519+224 (RBS~365)
and  J07472+503 (2MASS~J07471385+5020386) in Figs.~\ref{variabilityRBS365} and \ref{variabilityJ07472}.
In RBS~365, the core of the H$\alpha$ line taken in flare spectrum no. 1 shows an average amplitude, accompanied by 
a strong blue wing enhancement, extending to velocities of about $-450$~km\,s$^{-1}$
with a peak at $-86$~km\,s$^{-1}$. This is the only case in which we found signatures of material
close to the escape velocity of the star, which is about $500$~km\,s$^{-1}$ for mid- to early-type M~dwarfs.
We speculate that this integration mainly covers the quiescent state and ends 
right after flare onset; this can explain the presence of evaporating material without
associated core enhancement because
there is already heated material but not enough to show up against the
strong average H$\alpha$ core emission.
The associated \heir\ line profile in RBS~365 is deeper than the quiescent line profile.
This is consistent with elevated EUV irradiation at flare onset, which causes stronger
\heir\ line absorption. 
The \hei\ D$_{3}$ line shows no notable change.
This is in contrast to the other spectra of this star, where a
higher amplitude in H$\alpha$ is seen in combination with some fill-in of the \heir\ line
and additional emission in the \hei\ D$_{3}$ line; we show the spectrum no. 2
in Fig. \ref{variabilityRBS365} as an example.

One more intriguing example of a flare onset was observed in J07472+503 (2MASS~J07471385+5020386),
whose spectra
are shown in Fig.~\ref{variabilityJ07472}. The H$\alpha$ spectrum of spectrum no. 3 shows a prominent blue
asymmetry. Again this coincides with a deepening of the
\heir\ line that is likely attributable to increased EUV surface illumination,
but this line also shows a blue emission wing. The H$\alpha$ and \heir\ line wings both extend to velocities
of about $-200$~km\,s$^{-1}$ and peak around $-65$~km\,s$^{-1}$.
Additionally, we noticed a weak blue wing in the \hei\ D$_{3}$ line.
This implies that 
the densities in the moving material are high enough to drive the line into emission via collisions.
The flare spectrum no. 1 in Fig. \ref{variabilityJ07472} shows 
H$\alpha$ core enhancement without wing emission and also
seems to cover a flare. The response of the \heir\ line is 
a moderate fill-in. Curiously,
the largest pEW(\heir) is found for the spectrum no. 2, where
even an enhancement in the bluest and weakest component of the \heir\ triplet at about 10832~\AA\, can
be distinguished. The corresponding H$\alpha$ line profile looks 
completely innocuous, showing, if anything, a marginal blue wing.
We speculate that the \heir\ line reacts to an enhanced EUV radiation level
even before H$\alpha$ shows a significant response.
This is especially important for exoplanet studies of the \heir\ line, since it indicates that the
reaction of the stellar \heir\ line may precede that of H$\alpha$ in flares.

Although two more stars (J06574+740 (2MASS~J06572616+7405265)  and J06000+027 (G~099-049))
in the sample show a deepening of the \heir\ line with no significant change
seen in H$\alpha$, 
 the opposite can be found viz. a significant blue asymmetry in H$\alpha$ with no deepening
of the \heir\ line. This is seen for example for the star J05084-210 (2MASS~J05082729-2101444). 
The corresponding flares
possibly show lower EUV emission levels during onset or the \heir\ line response cannot be observed, for example
for a flare at or beyond the limb. 
Another rarely occurring case seems to be the connection between the deepening in the \heir\ line and a red
asymmetry in H$\alpha$. We could identify only a single spectrum (belonging to J22468+443, hereafter EV~Lac)
showing a deepening in the \heir\ line along with a minor H$\alpha$ red
asymmetry.

\subsection{Red asymmetries in H$\alpha$}

We identified a number of spectra showing red H$\alpha$ asymmetries
presumably caused by coronal rain in the later stages of flares \citep{asym}. 
Red H$\alpha$ asymmetries seem to be typically accompanied by
emission in the \heir\ line. 
This indicates that the flaring material
is dense enough to let collisions drive the \heir\ line into emission. 

\begin{figure*}
\begin{center}
\includegraphics[width=0.5\textwidth, clip]{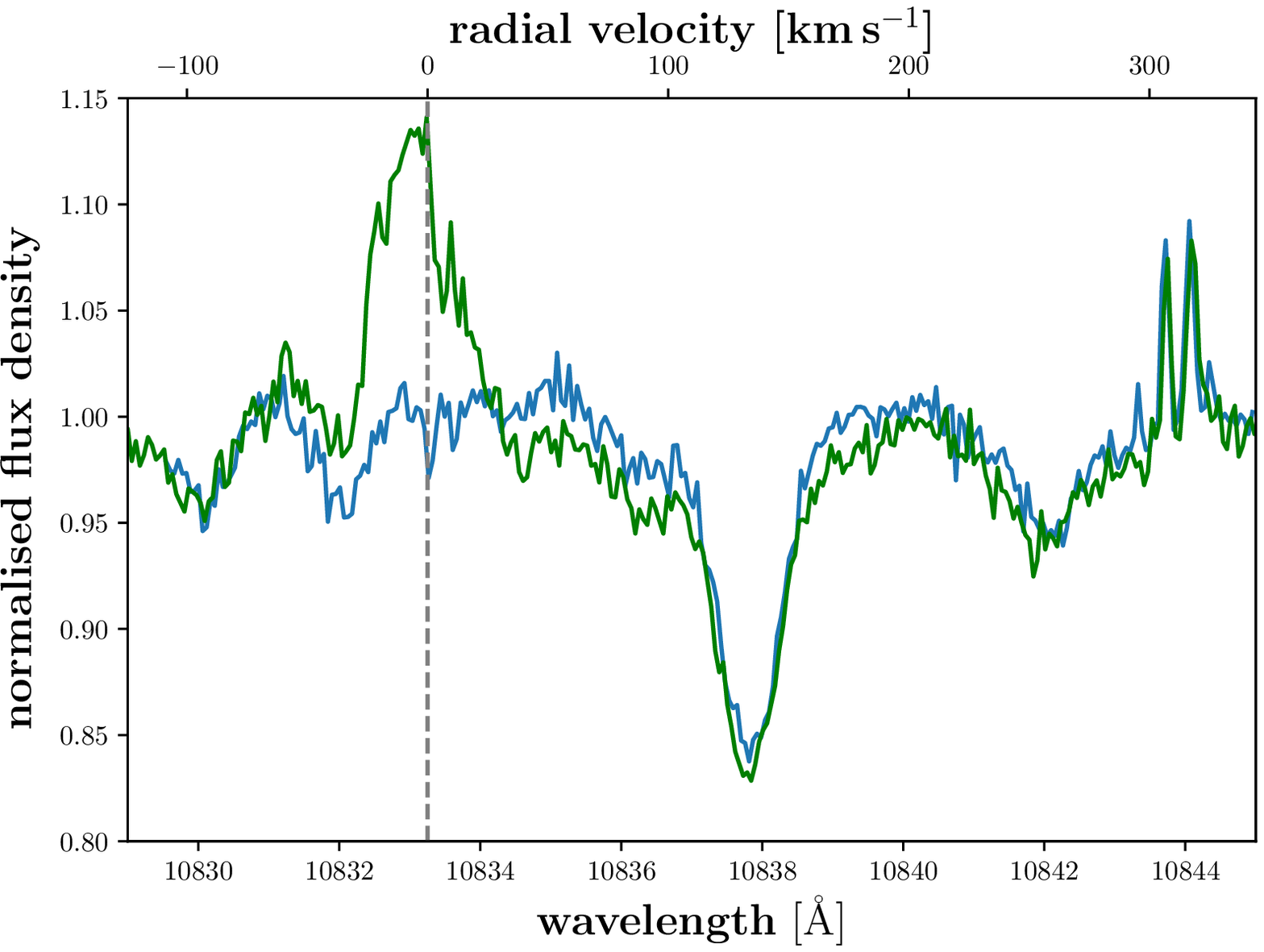}
\includegraphics[width=0.5\textwidth, clip]{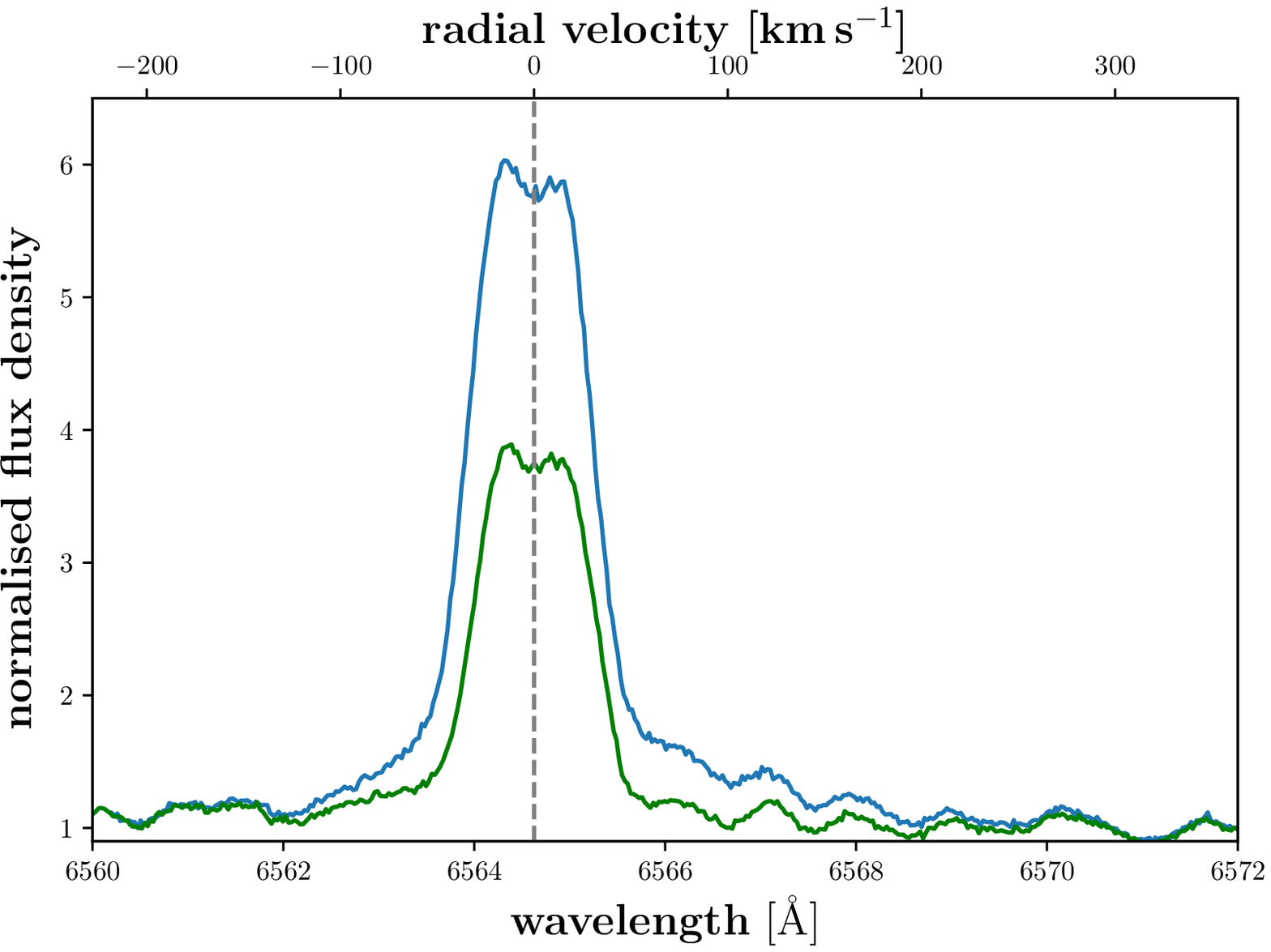}\\
\caption{\label{variabilityJ22518} Two of 11 spectra of GT~Peg.
  \emph{Left:} The \hei\ IR line. \emph{Right:} The H$\alpha$
  line. The green spectrum demonstrates the quiescent spectrum of the star.
}
\end{center}
\end{figure*}

As an example with emission in the main \hei\ IR line component but absorption in the wing,
 in Fig.~\ref{variabilityJ22518} we show
2 out of 11 spectra of J22518+317 (GT~Peg). One 
  spectrum demonstrates the typical quiescent chromospheric emission lines
  the other shows a flare spectrum with
an H$\alpha$ red asymmetry. In the flaring spectrum, the \heir\ line shows a prominent and slightly blue-shifted emission peak and 
a shallow, broad absorption feature on the red side of the line. This component is so broad that
it extends beyond the photospheric \ion{Na}{i} line at 10\,837.814~\AA\ on the red side.
The width of the line may be caused by turbulent broadening, but we deem it more probable that we are seeing
material with a range of velocities blending into one broad line since the line extends on the red side
to velocities of about 250-300~km\,s$^{-1}$ for both the \heir\ line and H$\alpha$ line. In this picture,
the down-raining material produces H$\alpha$ emission, while it is seen in absorption in the \heir\ line.
This suggests intermediate densities in the moving material,
which are too low to drive the \heir\ line into emission by collisions but
sufficient to produce absorption. 
Red wings can be identified neither in the \hei\, D$_{3}$
nor in the \ion{Ca}{ii} IRT lines.

The blue-shifted emission component in the \hei\ IR line
is also found in the \hei\, D$_{3}$ and in the \ion{Ca}{ii} IRT lines. The velocity
shift is about $-7$~km\,s$^{-1}$ for each of these lines. In the H$\alpha$ line, this blue-shifted component
is much less pronounced, producing only a slight
enhancement of the blue line flank. We speculate that this emission component originates from the flaring site
and its velocity shift corresponds to the rotational velocity of the flaring region.

\subsection{Symmetric line broadening of H$\alpha$}

\begin{figure*}
\begin{center}
\includegraphics[width=0.5\textwidth, clip]{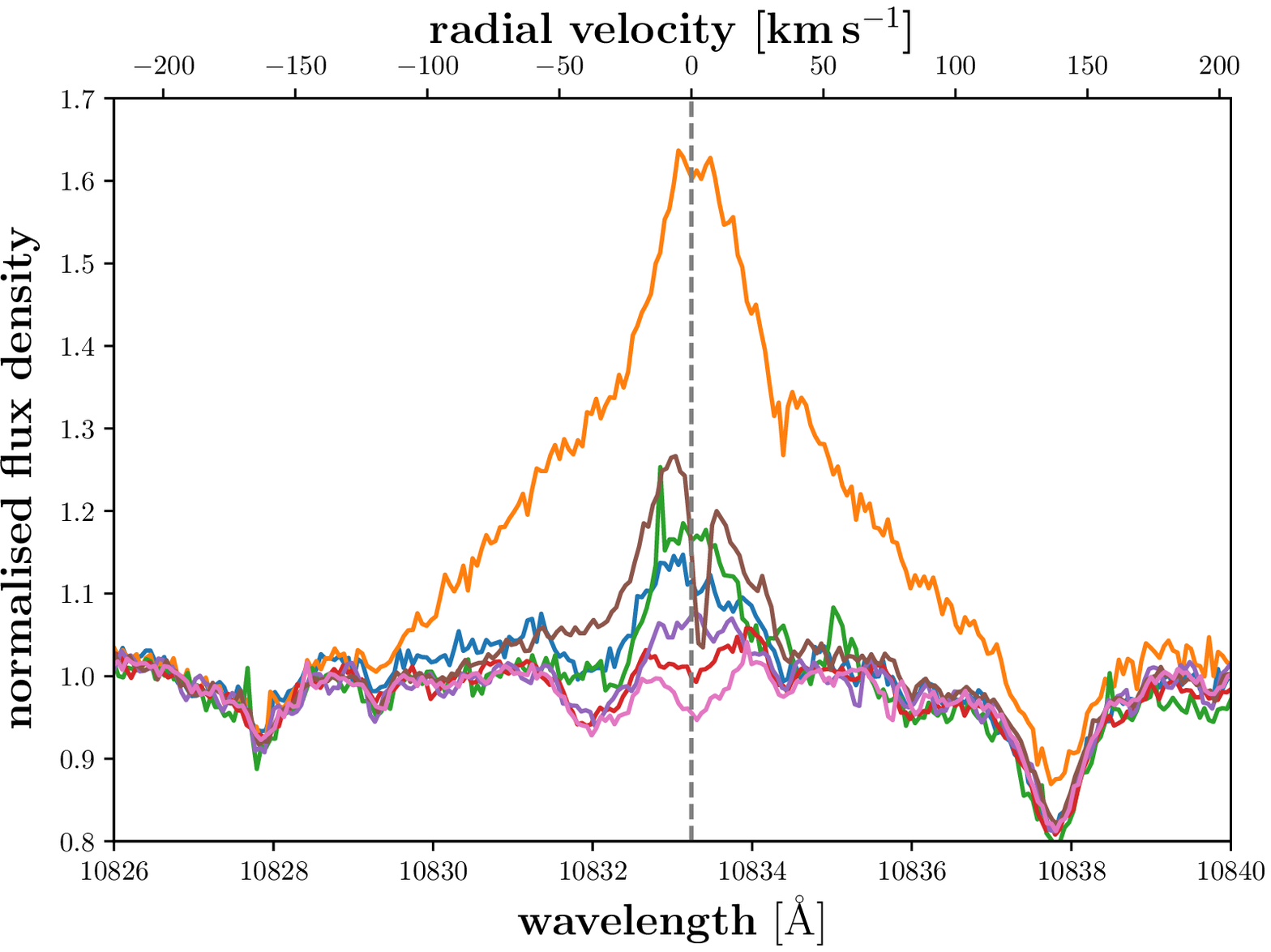}
\includegraphics[width=0.5\textwidth, clip]{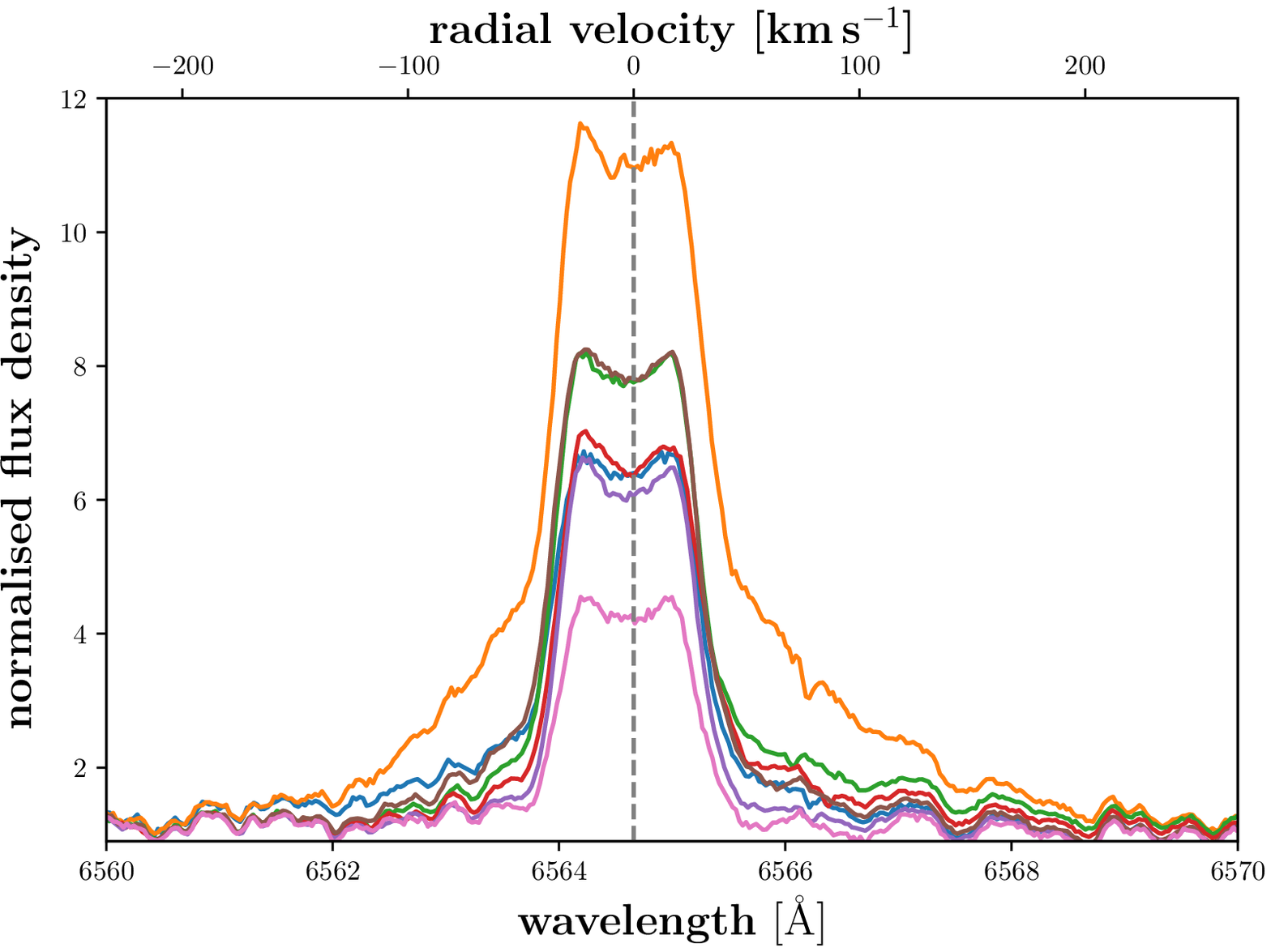}\\
\caption{\label{varevlacflareact} Example spectra of the flaring activity state of EV~Lac.
  \emph{Left:} The \hei\ IR line going from nearly continuum level into clear emission. \emph{Right:} The H$\alpha$
  line. For both panels the colours correspond to the same observation times as described in Table
    \ref{dateofflares}. The pink spectrum represents the quiescent state of EV~Lac.
}
\end{center}
\end{figure*}

We found a number of examples for
symmetric H$\alpha$ line broadening, and all of these go along with \heir\ line emission.
In Fig. \ref{varevlacflareact}, we show six spectra of EV~Lac, displaying
cases that we consider symmetric H$\alpha$ line broadening. 
The spectrum with the largest H$\alpha$ enhancement by far  (spectrum no. 2)
also shows \heir\ and \hei\,D$_{3}$ line emission with broad wings.
Apart from spectrum no. 6,
the other spectra also display approximately symmetric broadening.
Nevertheless, the symmetry is not perfect as, for example in spectra nos. 1 and 3. It remains
unclear whether this is attributable to physical line asymmetries.
We speculate that at least some of these line profiles are caused by 
integration effects.

Another example, for which we identified a
case of highly symmetric line broadening during a flare, is the star
J11474+667 (1RXS~J114728.8+664405).
 This star shows no detectable \heir\ IR line during the quiescent state,
but displays large symmetric wings in the H$\alpha$, \hei\ D$_{3}$, and \heir\ lines during a large flare.
Curiously, nothing similar is seen in the \ion{Ca}{ii} IRT lines.
The H$\alpha$ line wings extend to about $\pm400-450$~km\,s$^{-1}$. The
\heir\ line wings span the $\pm 100$~km\,s$^{-1}$ range and that of
the \hei\ D$_{3}$ line still reaches
$\pm 50-70$~km\,s$^{-1}$.
The  highest velocity material only seems to be bright in H$\alpha$, which may be attributable to low densities or a temperature favouring H$\alpha$ emission. This may also be
caused at least partially by the Stark effect, which affects the Balmer lines most strongly and is consistent with the non-detection of any wings in the 
\ion{Ca}{ii} IRT lines, which should not be affected by Stark broadening. 
Stark broadening could also
explain the extreme width of the line since material at such high velocities is
rarely observed and the velocities are too high to be caused by Alf\'enic turbulence \citep{Matthews, Lacatus},
which typically causes line widths of the order of 100~km\,s$^{-1}$. 
The \heir\ line normalised amplitude of this flare is among
the largest of the whole sample.


\begin{figure*}
\begin{center}
\includegraphics[width=0.5\textwidth, clip]{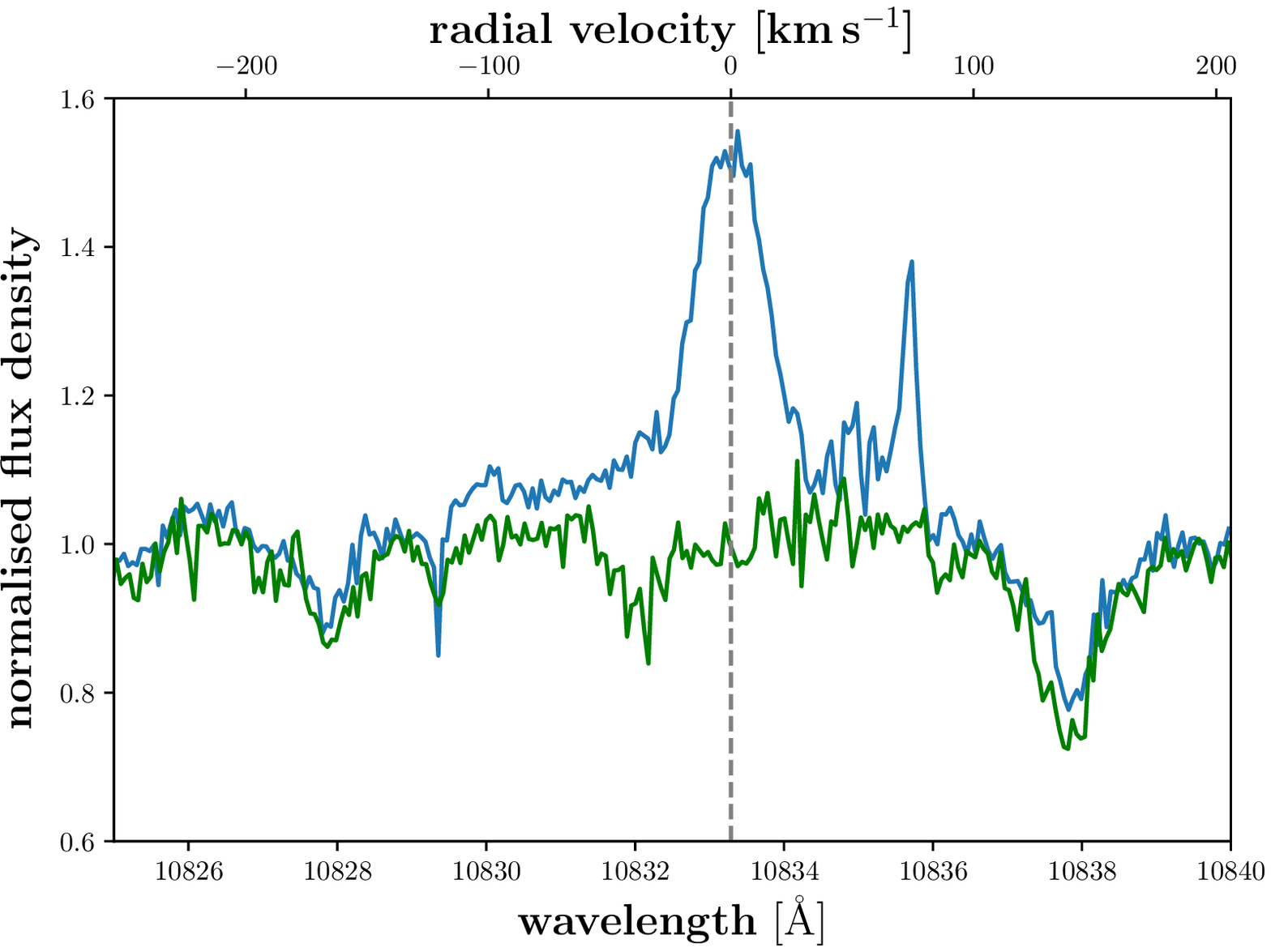}
\includegraphics[width=0.5\textwidth, clip]{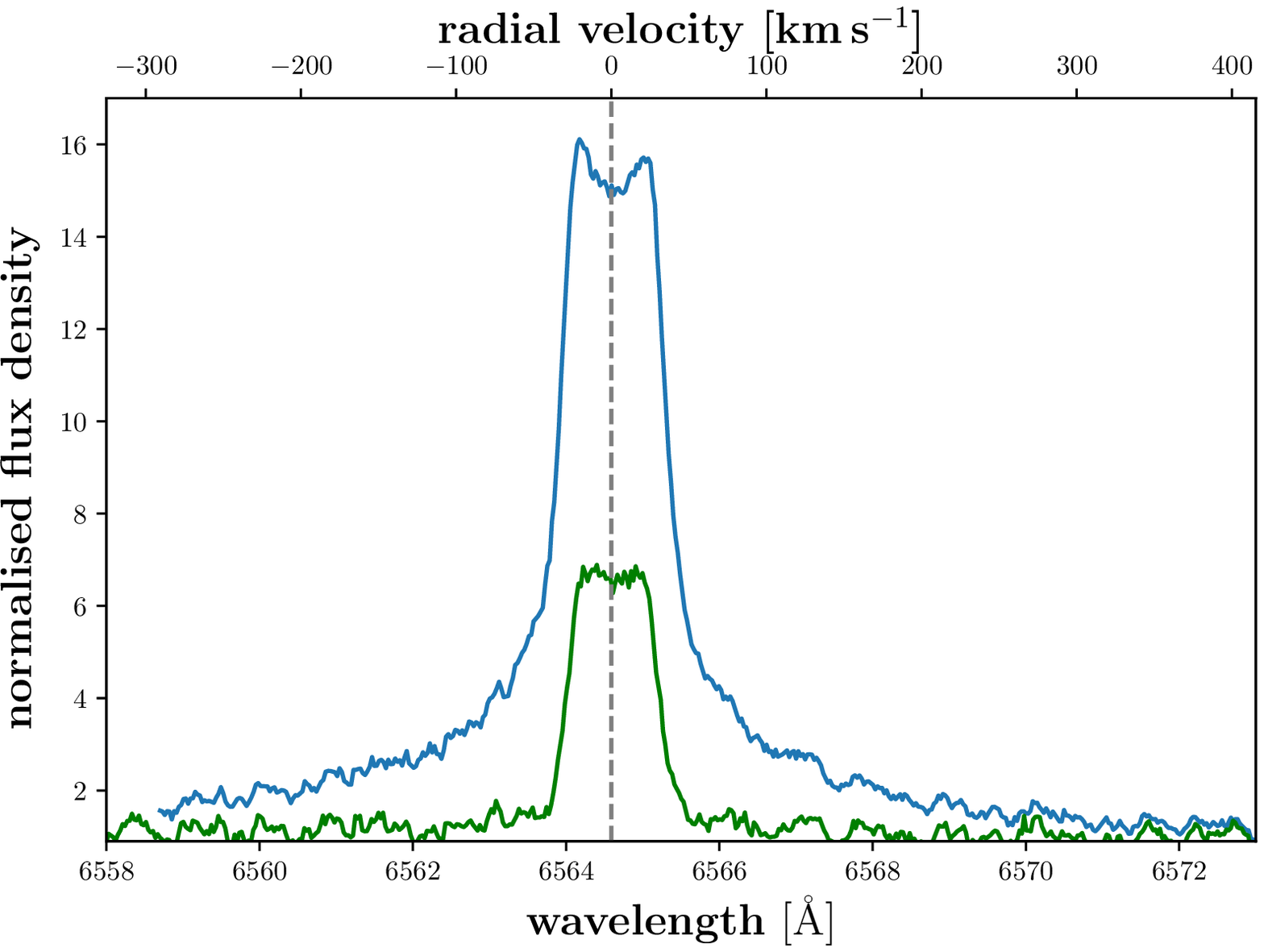}\\
\caption{\label{variabilityJ11474} Two of 21 available spectra for different chromospheric indicator
  lines for 
  J11474+667. \emph{Left:} The \hei\ IR line. \emph{Right:} The H$\alpha$
  line. The green spectrum illustrates the quiescent state. 
}
\end{center}
\end{figure*}

\subsection{LP~205-044}

\begin{figure*}
\begin{center}
\includegraphics[width=0.5\textwidth, clip]{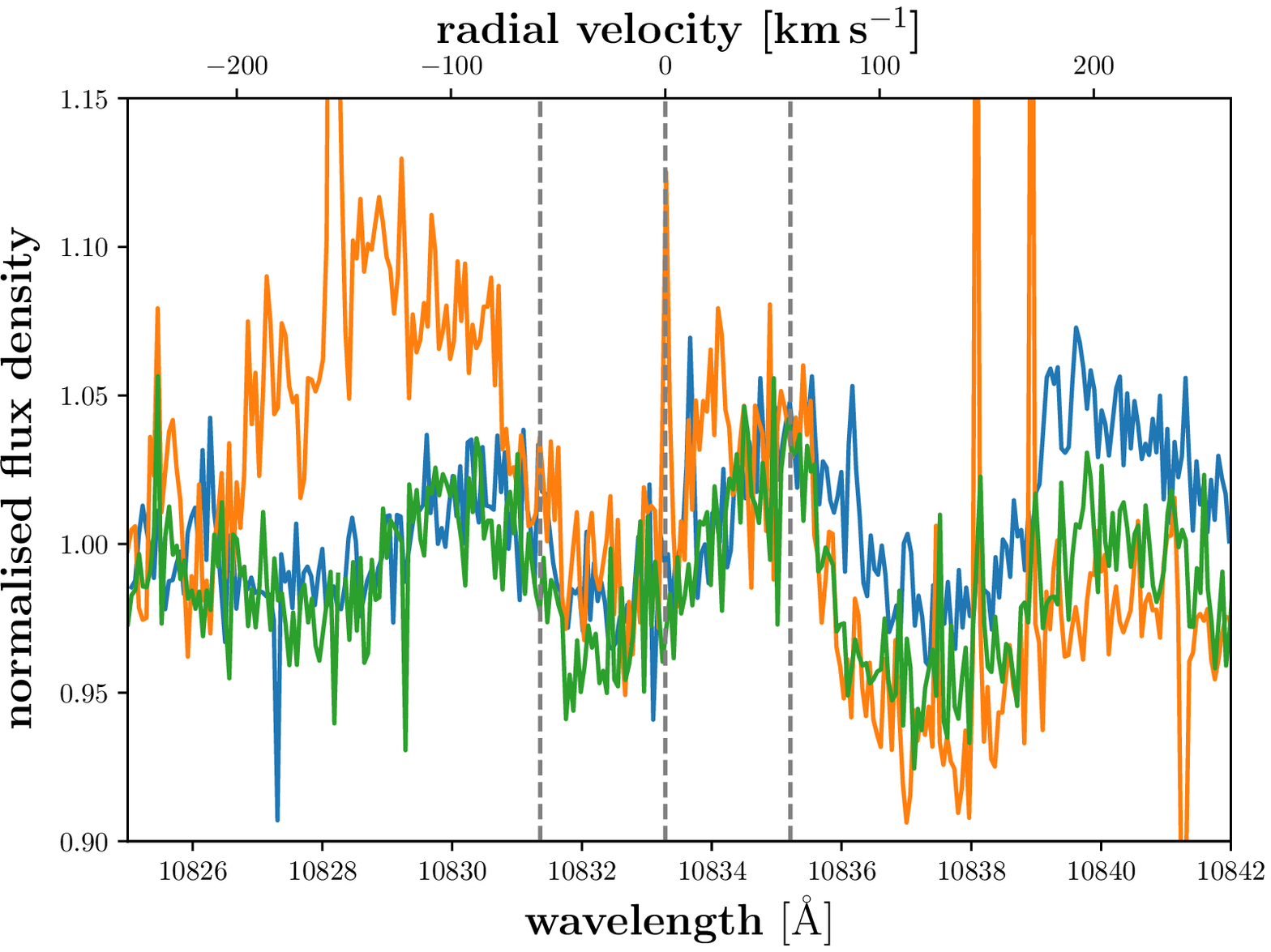}
\includegraphics[width=0.5\textwidth, clip]{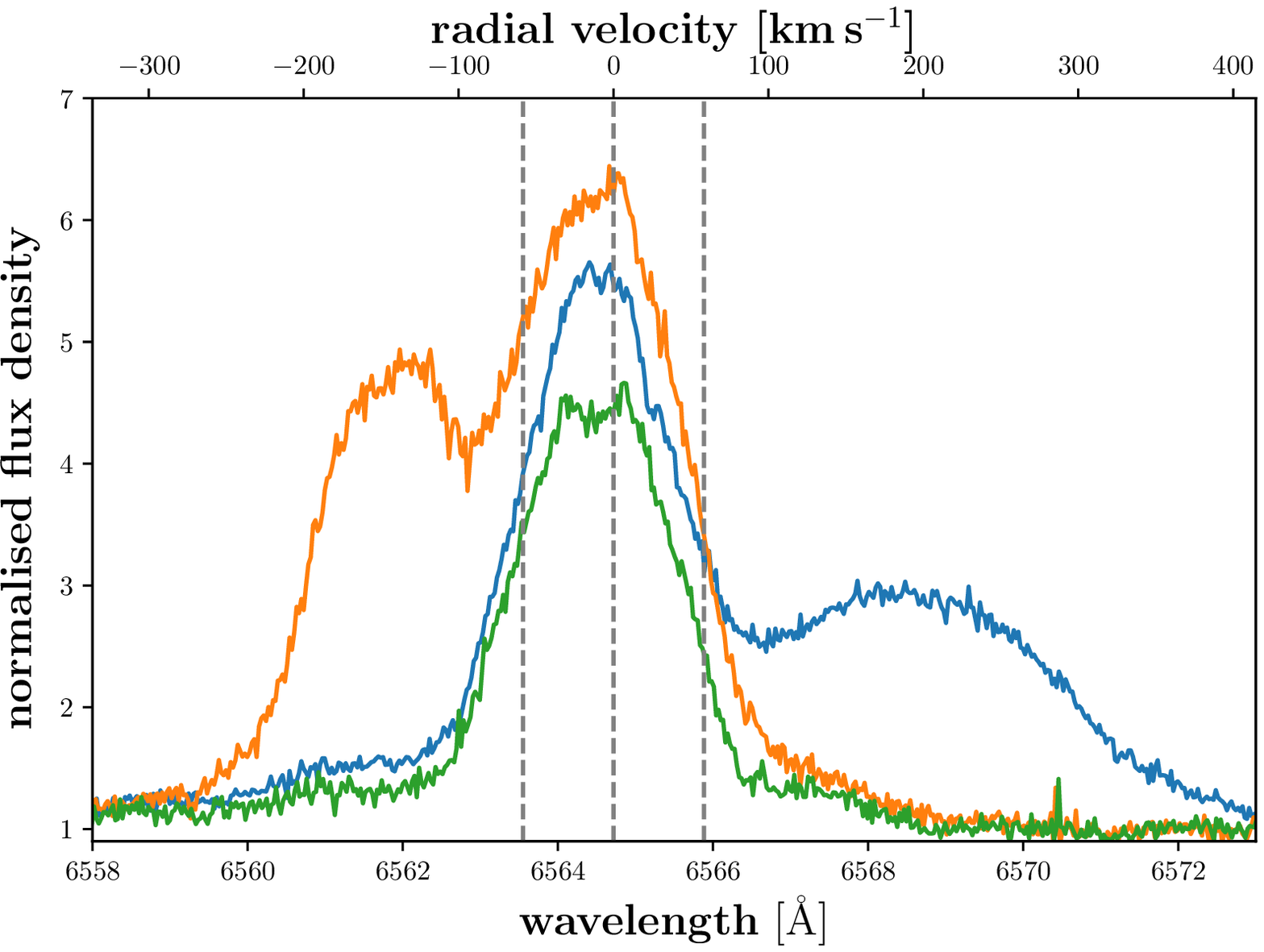}\\
\includegraphics[width=0.5\textwidth, clip]{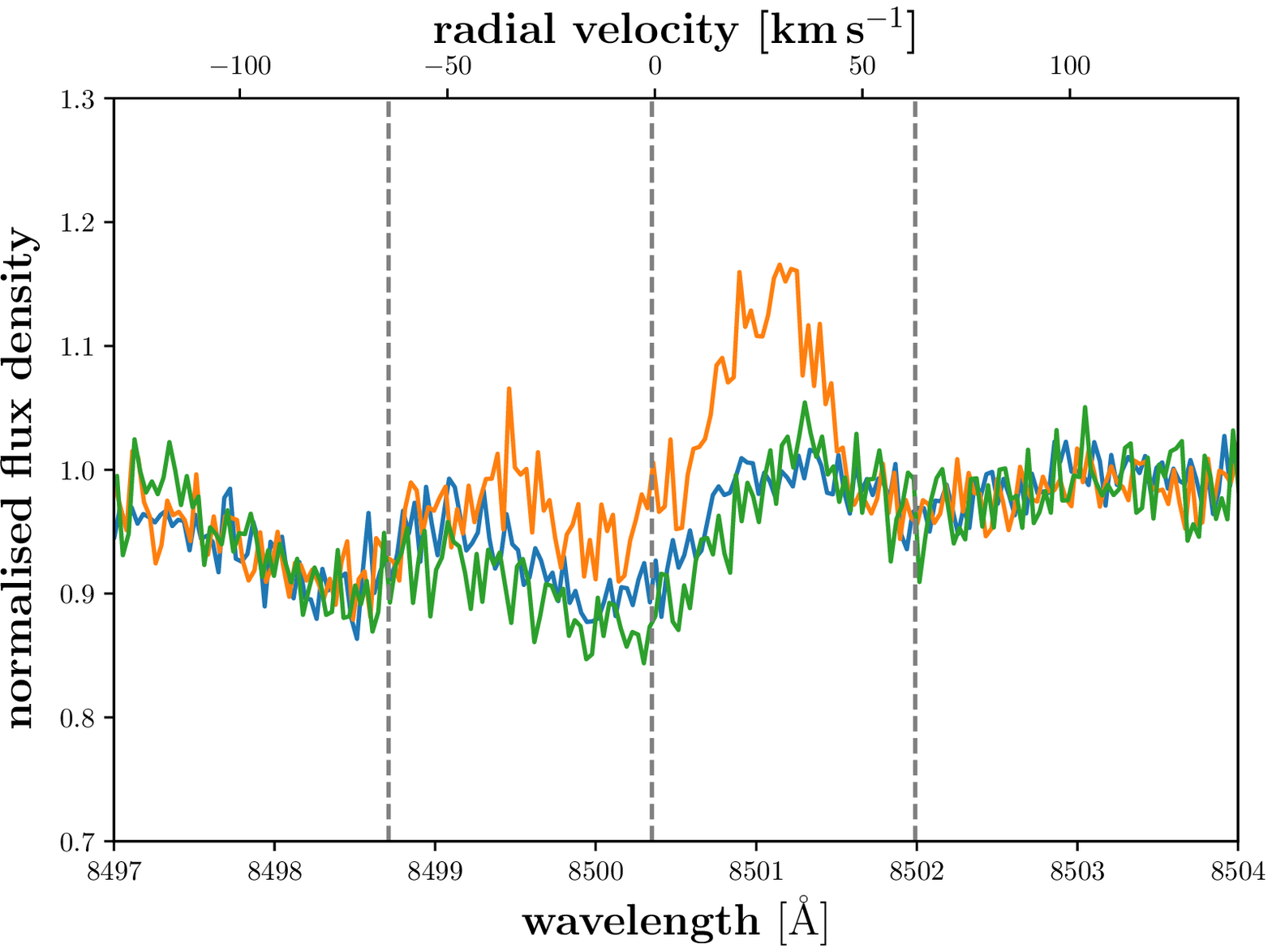}
\includegraphics[width=0.5\textwidth, clip]{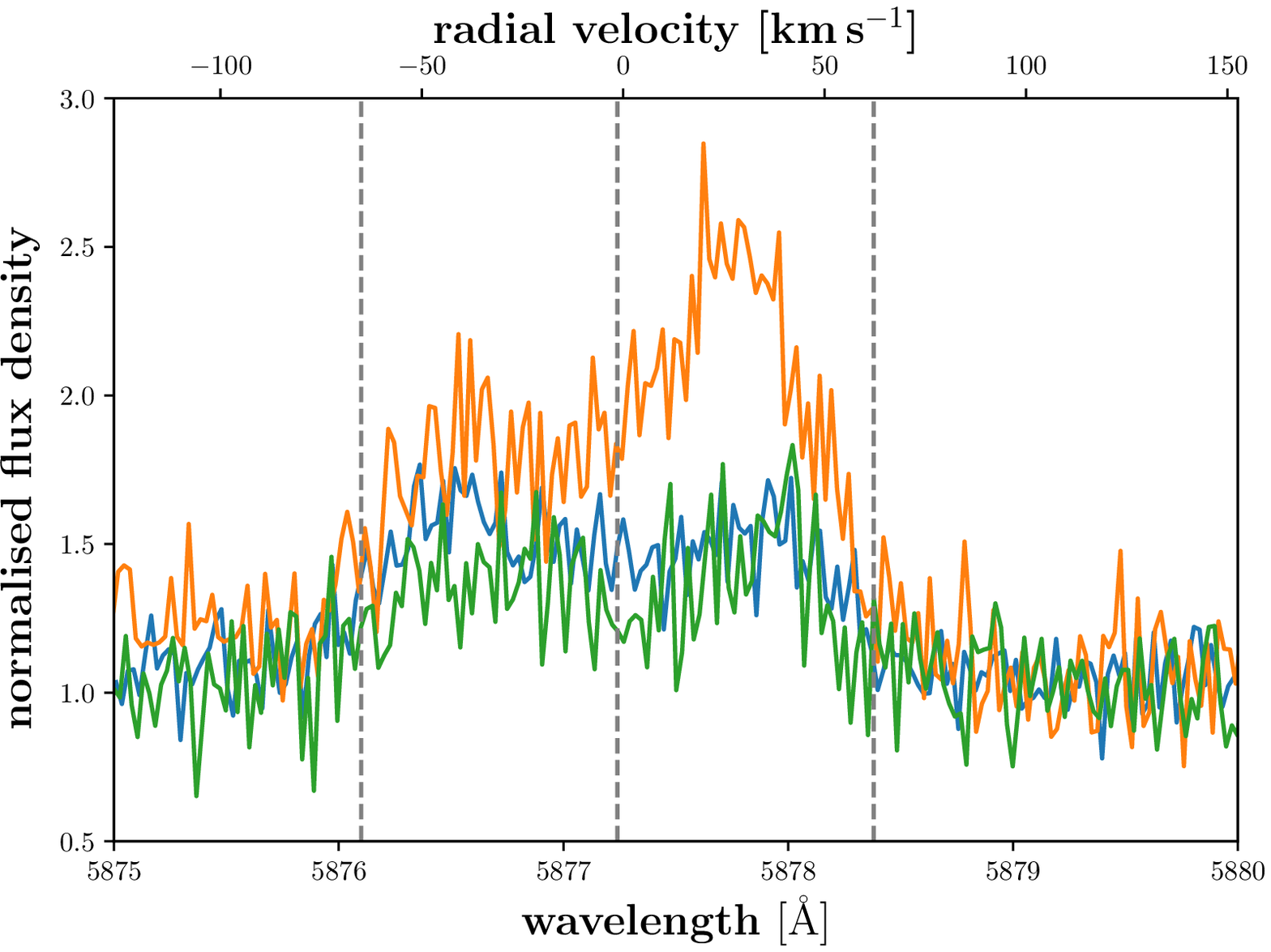}

\caption{\label{variabilityLP205-044}Three of 22 available spectra for different chromospheric indicator
  lines for 
  LP~205-044. \emph{Top left:} The \hei\ IR line. \emph{Top right:} The H$\alpha$
  line. \emph{Bottom left:} The \hei\ D$_{3}$ line. \emph{Bottom right:} The bluest \ion{Ca}{ii} IRT line. 
  The dashed vertical lines indicate\ the central wavelength of the respective chromospheric lines and the
  wavelengths corresponding to the maximum rotational velocity of \vsini = 58.4 km\,s$^{-1}$. The quiescent
activity level is represented by the green spectra. The blue spectrum no. 1 is taken
at JD 7823.44 and the orange spectrum no. 2 is taken at JD 8057.63 d.}
\end{center}
\end{figure*}

As one of the most outstanding examples of red and blue asymmetries, we show in
Fig.~\ref{variabilityLP205-044} the spectra of the
M5.0 star LP~205-044. Most of the spectra exhibit only moderate variation in the
H$\alpha$ line, which may be attributed to a quasi-quiescent state. However,
two spectra with extreme H$\alpha$
wings stand out. In both cases, the core of the H$\alpha$ line reacts only weakly.

The spectrum no. 2 shows an extreme blue wing in the H$\alpha$ line. A similar component
can be distinguished in the \heir\ line and, to a lesser extent,
in the \hei\ D$_{3}$ line. The mean shift of this component is about $-120$~km\,s$^{-1}$.
The \hei\ D$_{3}$ and the \ion{Ca}{ii} IRT lines simultaneously show  
a pronounced emission component red-shifted by about +30~km\,s$^{-1}$, that is within
the rotational line profile. We speculate that this red-shifted component originates in
the flare foot points, while the line wings trace evaporating material observed
during the onset of a flare. The origin of
a less strong blue-shifted component within the rotational profile of the \hei\ D$_{3}$
and the \ion{Ca}{ii} IRT lines remains unclear.

In contrast, the spectrum no. 2 in Fig.~\ref{variabilityLP205-044} shows an enormous
red H$\alpha$ wing with a mean shift of
about 180~km\,s$^{-1}$. The corresponding \heir\ line spectrum shows a less prominent broad enhancement
at a comparable shift. However, no such line enhancement is detectable in the \hei\ D$_{3}$
or \ion{Ca}{ii} IRT lines. We hypothesise that this observation reflects coronal rain observed towards
the end of a flare.

\subsection{Barta~161~12}

\begin{figure*}
\begin{center}
\includegraphics[width=0.5\textwidth, clip]{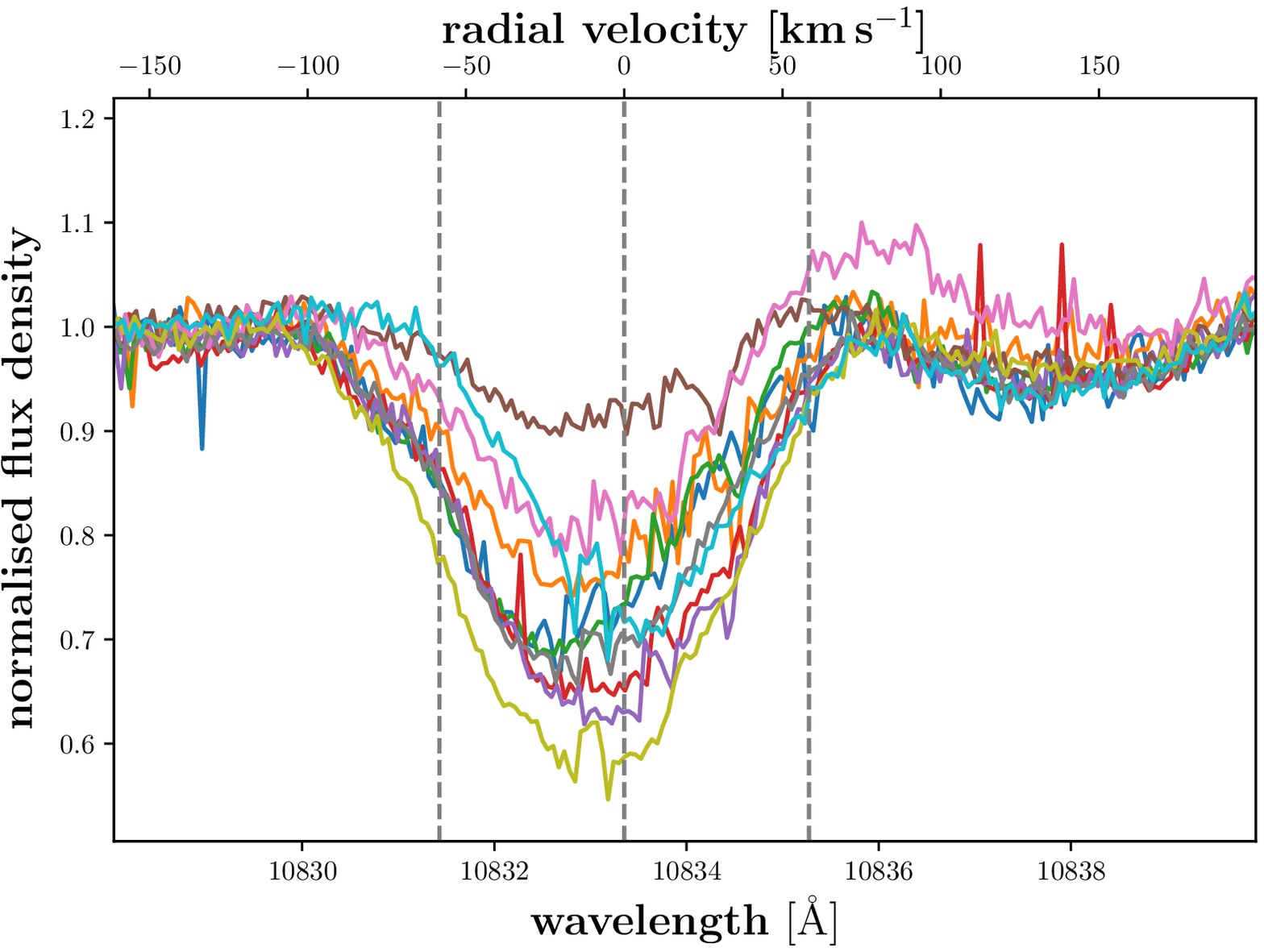}
\includegraphics[width=0.5\textwidth, clip]{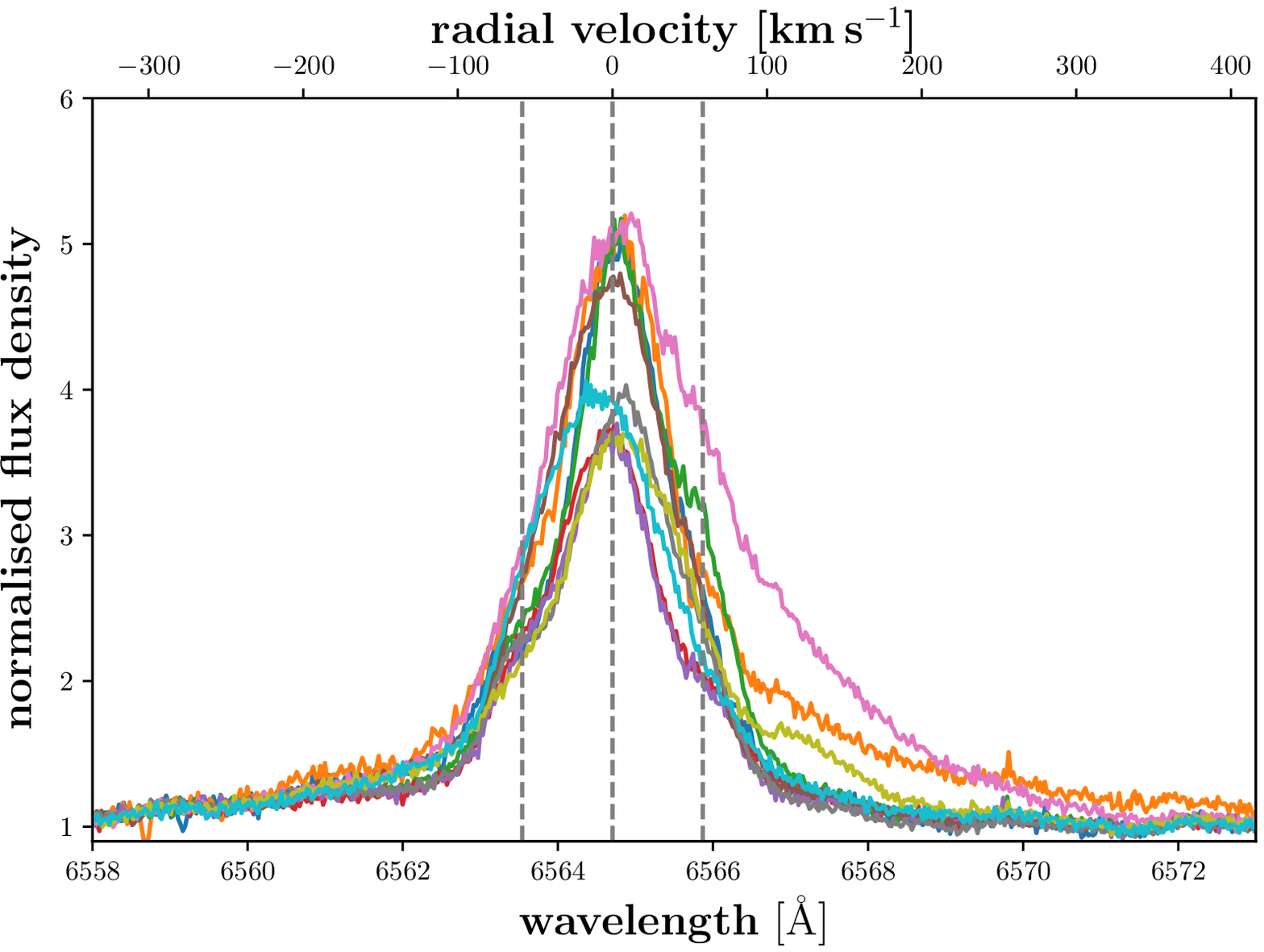}\\
\includegraphics[width=0.5\textwidth, clip]{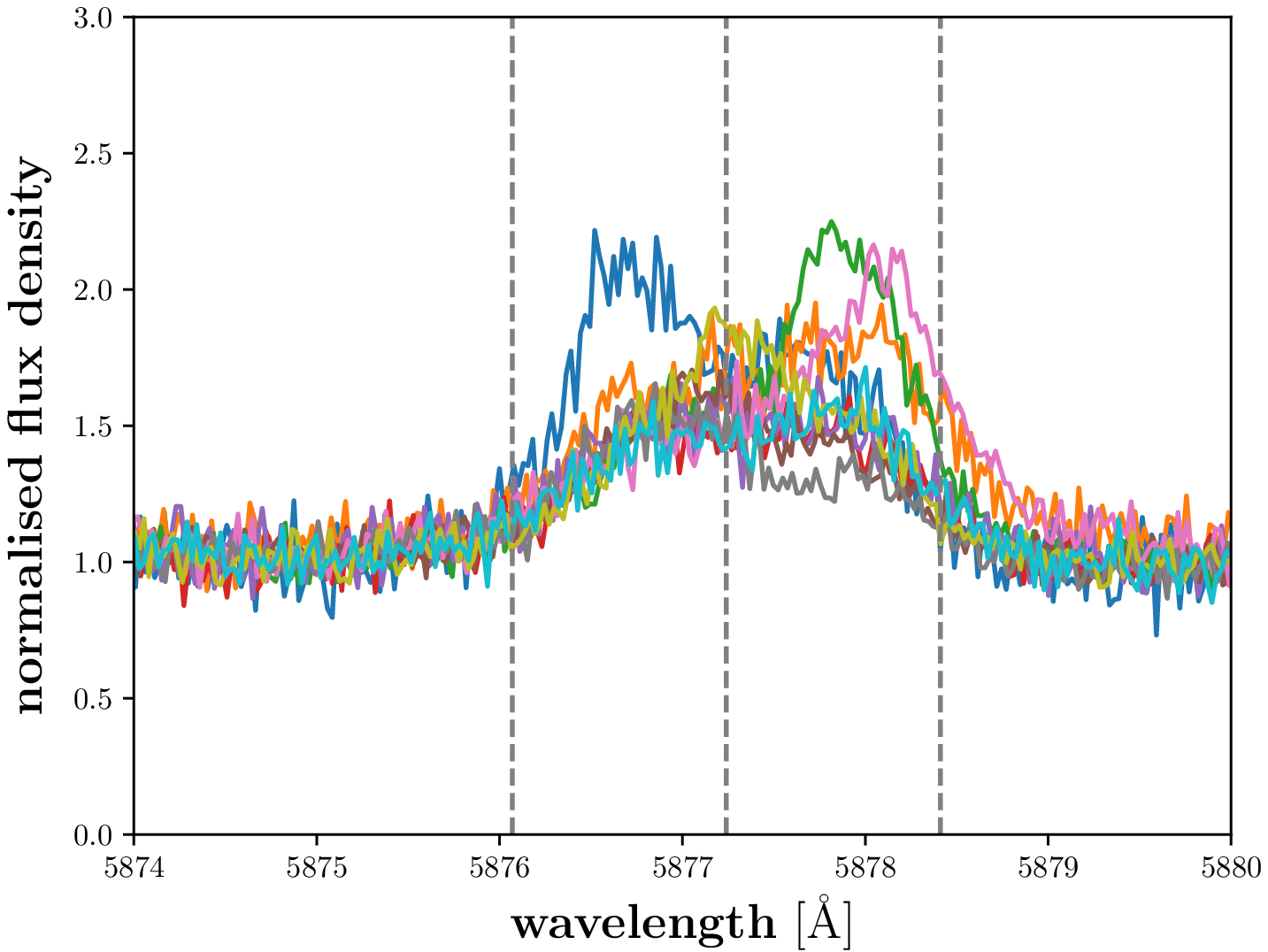}
\includegraphics[width=0.5\textwidth, clip]{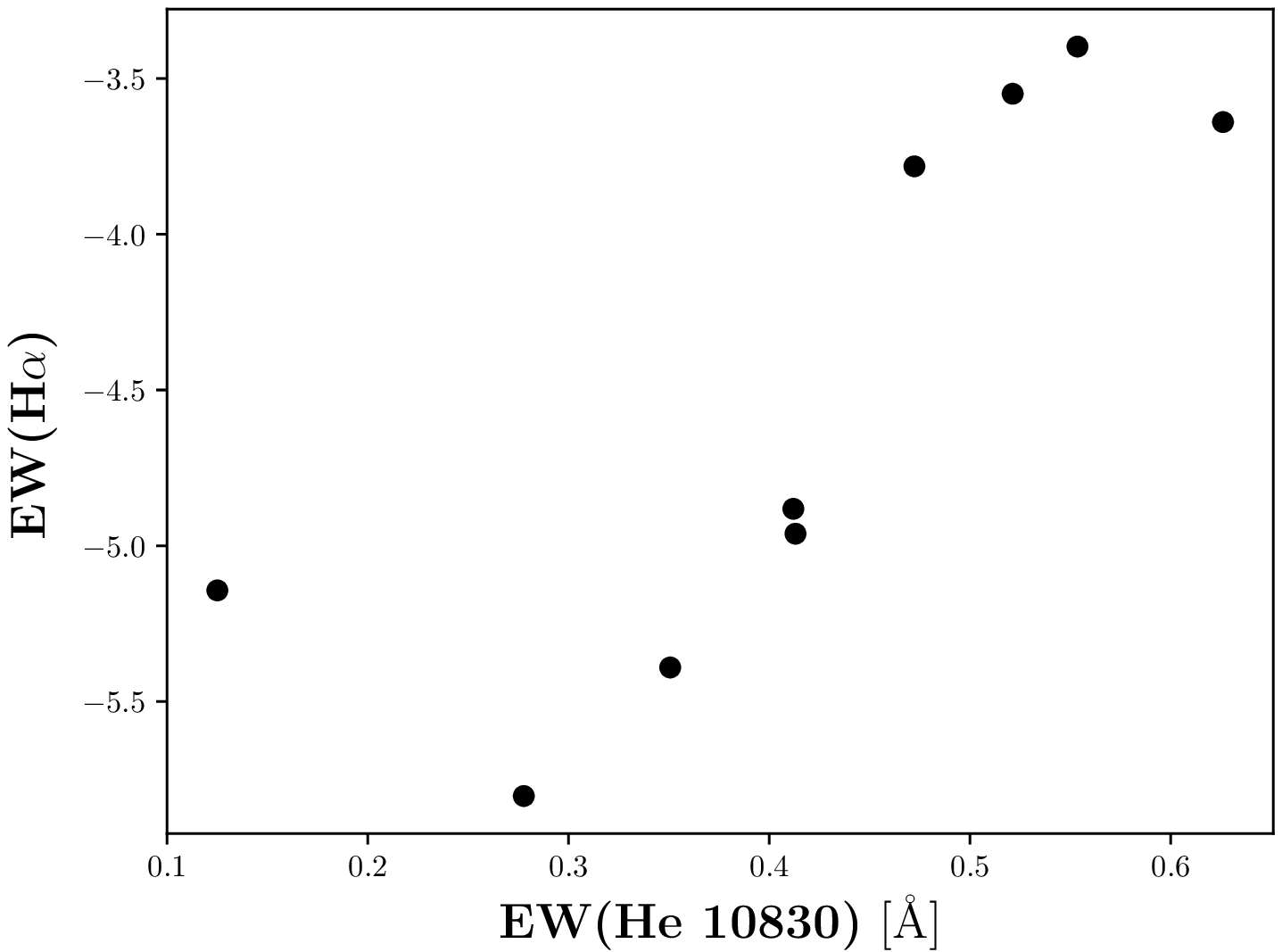}
\caption{\label{variabilityBarta} 
  Individual spectra for different chromospheric indicator lines for the M4.0 star
  Barta~161~12. \emph{Top left:} The \hei\ IR line. \emph{Top right:} The H$\alpha$
  line. \emph{Bottom left:} The \hei\ D$_{3}$ line. \emph{Bottom right:} Correlation between pEW(\hei)
and pEW(H$\alpha$).}
\end{center}
\end{figure*}


Barta~161~12 is a fast rotator with \vsini\ of 59.8~km\,s$^{-1}$ \citep{Reiners2017}
and a photometrically derived rotation period of 0.7~d \citep{Kiraga2012}.
As a candidate member of the
$\beta$\,Pic moving group, the star is young with an estimated age of 12-22~Myr \citep{Malo2013}.
The strong line broadening in this star leads to ubiquitous blends. In particular,
the \heir\ line is blended with the neighbouring, 
unidentified lines.

In Fig.~\ref{variabilityBarta}, we show all ten \heir, H$\alpha$, and \hei\ D$_{3}$ spectra
of Barta~161~12. The \heir\ line  especially shows an overall degree of variability, which is unmatched
by any other target in our sample. By inspection of the \hei\ D$_{3}$ line spectra, we
identify four spectra with clear emission components (orange, magenta, green, and blue),
which we attribute to flaring. However, also for the remaining spectra, no clear quiescent
state can be identified in this star. This suggests that flaring is not the sole source of
modulation of the chromospheric line profiles in
Barta~161~12. 
Comparison of the pEWs of the H$\alpha$ and \heir\ lines shows a good correlation with elevated
H$\alpha$ emission being associated with fill-in of the \heir\ line.
A potential source of the observed variability is rotational modulation. To test this
hypothesis, however, short-cadence spectroscopy, sampling the rotation period would be required.
Its high level of overall variability makes Barta~161~12 a highly promising target for
short-cadence follow-up.

\section{Implications for exoplanet atmosphere studies}
For more than half of the stars with \hei\ line variability, the line pEW is also correlated to pEW(H$\alpha$).
Low correlation coefficients for these stars are normally caused by outliers owing to telluric artefacts, noise,  or
flaring activity. On the other hand, for non-variable stars the pEW(\hei) is indeed  constant even if 
some variation in pEW(H$\alpha$) or pEW(\cab) is present.
This implies a comparable
insensitivity of the stellar \heir\ line to activity phenomena on M~dwarfs, which is
a favourable property for planetary atmospheric studies, for example by \heir\ line
transmission spectroscopy. The generally good relation between the pEW changes observed in
the \heir\ line with other activity indicators furthermore allows us
to identify \heir\ line variability by monitoring more sensitive
chromospheric lines, such as the H$\alpha$ or \ion{Ca}{ii} H \& K lines, if available in the spectrum, which we strongly recommend.

While choosing inactive host stars is an obvious route to minimise activity-related inference
in planetary atmospheric studies, the peculiar formation of the \heir\ line makes
higher host star activity levels
desirable for studies of planetary \heir\ lines \citep[e.g.][]{Nortmann2018,Salz2018}. Fortunately,
our study suggests that variations in the stellar \heir\ line remains
comparably insensitive and a non-variable H$\alpha$ line also implies a stable \heir\ line.
The relation may break down in strong flares, which are however recognised with little
difficulty in virtually all chromospheric lines.

For searches of the \heir\ line in planetary atmospheres our outstanding examples are also instructive. The snapshot
character of our data prevents us from estimating timescales on which the stellar \heir\ line
is variable and from comparing these variability timescales to timescales of in-transit phases of exoplanets. Nevertheless, the possibly strong
reaction of the stellar \heir\ line to especially larger flares shows that timescales of variation can be
in the range of minutes, again showing the need to monitor the stellar chromospheric variability using
other chromospheric lines.

Moreover, we studied the more than 100 spectra of EV~Lac, of which many are taken only a day apart. This
 active M3.5 dwarf reveals a very shallow line at best by visual inspection. 
Also, the \heir\ line seems to be  stable for much of the time,
with episodes of increased activity and a number of flares causing the variability.

\section{Conclusions}

We present a variability study of the \hei\ IR line as observed in 319 M dwarfs by inspecting 
more than 14\,000 CARMENES spectra. We find that the \hei\ IR line can be very stable especially for early M dwarfs
with H$\alpha$ in absorption.  Nevertheless, we find variability in 56 (18 \%) out of 319 stars in this line.
All of these stars show H$\alpha$ in emission, and if we examine only the sub-sample
of stars with \vsini > 25 km\,s$^{-1}$ we find about 80 percent
of these stars to be variable in the \hei\ IR line. Also mid-type M dwarfs, which were found in our
previous paper \citep{hepaper}
to display no \hei\ IR line in the average spectrum, exhibit some \hei\ line variability. For many
of these stars the \hei\ line occasionally turns from an absorption into an emission line, thus averaging out in the mean spectrum; this is, for example,  the case for G~080-021 in Fig. \ref{J03473}.
Alternatively, the star does not display a \hei\ line during the 
quiescent state, but does so during the flaring state, as shown, for
example, by J22518+317/GT~Peg in Fig. \ref{variabilityJ22518}. Most stars that are variable in the \hei\ line are
mid-type M3.0 to M6.0\,V stars, where a steep decline for the line detection or no
line detection at all is found for the average spectrum.

Generally, the \heir\ line correlates well with H$\alpha$ if it shows detectable variability. For stars whose \heir\ line shows no 
variability, H$\alpha$ may nevertheless do so, leading to non-correlation in these cases and stressing the relative insensitivity
of the \heir\ line to intrinsic variability; this makes the line a promising target for atmospheric studies of exoplanets using
transmission spectroscopy.

Moreover, we present a number of outstanding examples of
chromospheric line variations attributable to flares.
Flaring activity manifests itself in H$\alpha$ line broadening combined with possible
line asymmetries. A deepening of the \hei\ IR line is typically observed during flare
onset, where the H$\alpha$ line is still only marginally enhanced and often exhibits blue asymmetries as caused
by chromospheric evaporations, which are thought to
characterise the early stages of flare evolution. The deepening is likely produced by enhanced EUV
and X-ray irradiation levels of the stellar atmosphere caused by the flare
and therefore larger photon-ionisation and recombination rates of \ion{He}{i}.
In later flare stages, the \hei\ IR line tends to go into emission. All
H$\alpha$ spectra exhibiting symmetric broadening or red asymmetries also have \hei\ IR emission lines,
regardless of whether the line is observed in absorption or remains undetectable during the quiescent state of the star. 
Different chromospheric lines may behave similarly in terms of velocity shift, but
there can also be remarkable differences, which we cannot currently
explain. This may be caused by the snapshot character of our observations. 
Clearly, a detailed analysis of the \heir\ line variability on shorter flare and transit timescales
would greatly benefit from continuous, short-cadence time series observations. 

\begin{acknowledgements}
  B.~F. acknowledges funding by the DFG under Schm \mbox{1032/69-1}.
  CARMENES is an instrument for the Centro Astron\'omico Hispano-Alem\'an de
  Calar Alto (CAHA, Almer\'{\i}a, Spain). 
  CARMENES is funded by the German Max-Planck-Gesellschaft (MPG), 
  the Spanish Consejo Superior de Investigaciones Cient\'{\i}ficas (CSIC),
  the European Union through FEDER/ERF FICTS-2011-02 funds, 
  and the members of the CARMENES Consortium 
  (Max-Planck-Institut f\"ur Astronomie,
  Instituto de Astrof\'{\i}sica de Andaluc\'{\i}a,
  Landessternwarte K\"onigstuhl,
  Institut de Ci\`encies de l'Espai,
  Institut f\"ur Astrophysik G\"ottingen,
  Universidad Complutense de Madrid,
  Th\"uringer Landessternwarte Tautenburg,
  Instituto de Astrof\'{\i}sica de Canarias,
  Hamburger Sternwarte,
  Centro de Astrobiolog\'{\i}a and
  Centro Astron\'omico Hispano-Alem\'an), 
  with additional contributions by the Spanish Ministry of Economy, 
  the German Science Foundation through the Major Research Instrumentation 
    Programme and DFG Research Unit FOR2544 ``Blue Planets around Red Stars'', 
  the Klaus Tschira Stiftung, 
  the states of Baden-W\"urttemberg and Niedersachsen, 
  and by the Junta de Andaluc\'{\i}a.

\end{acknowledgements}

\bibliographystyle{aa}
\bibliography{papers}

\end{document}